\documentstyle[graphics,graphicx,times]{mn2e}
\input{psfig}

\newcommand{\transi}[2]{#1\textsc{#2}}

\def\EQN#1{\label{e:#1}}        

\def\Sec#1{{Section~\ref{s:#1}}}
\def\Tab#1{{Table~\ref{t:#1}}}        

\def\kms{\mbox{~km~s$^{-1}$}}
\def\hMpc{\mbox{$h^{-1}$Mpc}}
\def\hkpc{\mbox{$h^{-1}$kpc}}

\def\zem{$z_{\rm em}$}
\def\zabs{$z_{\rm abs}$}

\def\parn{\par\noindent}

\def\lya{\mbox{Lyman-$\alpha$}}

\def\Ol{\Omega_\Lambda}

\def\loss{lines of sight}

\def\cIV{\transi{C}{iv}~}

\begin{document}

\title[
Transverse correlation in the Lyman-$\alpha$ forest 
]{
Transverse and longitudinal correlation functions in the Intergalactic
Medium from 32 close pairs of 
high-redshift quasars\thanks{Based on observations carried out at the 
European Southern Observatory  with UVES (ESO program No. 65.O-299 and 
the Large Program 'The Cosmic Evolution of the IGM' No. 166.A-0106),
FORS2 (ESO program No. 66.A-0183) and FORS1 (ESO programs No. 69.A-0457 and 70.A-0032) 
on the 8.2~m VLT telescopes Antu, Kuyen and Melipal operated at Paranal 
Observatory; Chile.}
}

\def\inst#1{{${}^{#1}$}}

\author[Coppolani et al.]{F. Coppolani\inst{1}, P.~Petitjean\inst{2,3}, 
F. Stoehr\inst{2}, E.~Rollinde\inst{4,2},
C. Pichon\inst{2}, S. Colombi\inst{2}, 
\newauthor M.G.  Haehnelt\inst{5}, B. Carswell\inst{5}, R. Teyssier\inst{6} \\
$^1$ European Southern Observatory, Alonso de C\'ordova 3107, Casilla 19001, Vitacura, Santiago, Chile 
 - email: fcoppola@eso.org \\
$^2$ Institut d'Astrophysique de Paris, UMR 7095 CNRS \& Universit\'e Pierre et Marie
Curie, 98 bis boulevard d'Arago, 75014 Paris, France \\
$^3$ LERMA, Observatoire de Paris, 61, avenue de l'observatoire F-75014
        Paris, France \\
$^4$ IUCAA, Post Bag 4, Ganesh Khind, Pune 411 007, India\\
$^5$ Institute of Astronomy, Madingley Road, Cambridge CB3 0HA, UK\\
$^6$ DAPNIA, CEA Saclay, Bat 709, 91191 Gif-sur-Yvette France
}
\date{Typeset \today ; Received / Accepted}
\maketitle

\begin{abstract}

We present the transverse flux correlation function of  the 
Lyman-$\alpha$ forest in quasar absorption spectra at 
$z$~$\sim$~2.1 from  VLT-FORS and VLT-UVES observations of a total of 
32 pairs of quasars; 26 pairs with separations in the range 0.6~$<$~$\theta$~$<$~4~arcmin
and 6 pairs with 4~$<$~$\theta$~$<$~10~arcmin. Correlation is detected at the 
3~$\sigma$ level up to separations of the order of  $\sim$4~arcmin
(or $\sim$4.4$h^{-1}$~Mpc comoving at $z=2.1$ for $\Omega_{\rm m}$~=~0.3
and $\Omega_{\Lambda}$~=~0.7). We have, 
furthermore, measured the longitudinal correlation function at a somewhat higher
mean redshift ($z$~=~2.39) from  20 lines of sight
observed with high spectral resolution and high signal-to-noise ratio
with VLT-UVES. We compare the observed transverse and
longitudinal correlation functions 
to that obtained from numerical simulations and illustrate
the effect of spectral resolution, thermal broadening and peculiar
motions. 
The shape and correlation length of the correlation 
functions are in good agreement with those expected from absorption by the 
filamentary and sheet-like structures in the photoionized warm intergalactic medium 
predicted in CDM-like models for structures formation. 
Using a sample of 139 C~{\sc iv} systems detected along the lines of sight 
toward the pairs of quasars we also investigate the transverse correlation of metals
on the same scales. The observed transverse correlation function
of intervening C~{\sc iv} absorption systems is consistent with that of a randomly
distributed population of absorbers. This is likely due to the small number of pairs 
with separation less than 2 arcmin. We detect, however,  a significant overdensity of
systems in  the sightlines towards the  quartet Q~0103$-$294A\&B, Q~0102$-$2931 
and Q~0102$-$293 which extends over the redshift range
1.5~$\leq$~$z$~$\leq$~2.2  and an angular scale  larger than 
10~arcmin. 
\end{abstract}



\section{Introduction}

\parn 
The  numerous  H~{\sc i} absorption  lines seen  in the  spectra  
of  distant  quasars, the  so-called Lyman-$\alpha$ forest, contains
precious information on the spatial  distribution of neutral
hydrogen in the Universe. Unravelling this information from
individual spectra has for a long time proven difficult and ambiguous 
(see Rauch 1998 for a review). 
Studies of the correlation of the Lyman-$\alpha$ forests observed in the two spectra
of  QSO pairs have been instrumental in  measuring the spatial extent 
of absorbing structures. The Lyman-$\alpha$  forests in  the spectra of multiple 
images of lensed quasars or pairs of quasars with  separations of a few
arcsec (Bechtold et al. 1994: Dinshaw et al. 1994; Smette et al. 1995;
Impey et al. 1996; Rauch, Sargent \& Barlow 1999; Becker, Sargent \& Rauch 2004) 
appear nearly identical implying that the absorbing 
structures have sizes $>$50$h_{70}^{-1}$~kpc. Significant correlation 
between absorption spectra of adjacent lines of sight toward quasars 
still exists for separations of a few to ten arcmin suggesting 
a size or better a coherence length of the structures larger than
500~$h_{70}^{-1}$~kpc (e.g. Shaver \& Robertson 1983; Dinshaw et
al. 1997;  Petitjean et al. 1998; D'Odorico et al. 1998; Crotts \& Fang 1999; 
Young, Impey \& Foltz 2001; Aracil et al. 2002, Rollinde et al. 2003)  and a
non-spherical geometry of the  absorbing structures (Rauch \& Haehnelt 1995; Rauch et
al. 2005). On even larger scales,  Williger et al. (2000) 
still find evidence for an excess of clustering on 10~Mpc scales.
\parn
Numerical simulations of the warm photoionized Intergalactic Medium 
within the framework of cold dark matter (CDM) like models of
structure formation have demonstrated  that the neutral gas density 
traces the underlying  dark matter density field on scales larger than 
the Jeans length of the gas (e.g. Cen et  al. 1994;  Petitjean, M\"ucket \& 
Kates 1995; Theuns et al. 1998). The picture of 
the  Lyman-$\alpha$ forest arising from the density fluctuations in  
a warm photoionized Intergalactic Medium distributed as expected in a  
CDM  model explains the statistical properties of individual QSO 
absorption spectra very well (see Weinberg et al. 1999 for a review). 
Most of the baryons are located in filaments and
sheets which are only overdense by factors of a few and produce 
absorption in the column density range 
10$^{14}$~$<$~$N_{\rm HI}$~$<$~10$^{15}$~cm$^{-2}$ at $z\sim 2$. On the other hand, 
most of the volume is occupied by underdense regions 
that produce absorption with $N_{\rm HI}< 10^{14}$~cm$^{-2}$. 
Analytical calculations and numerical simulations of the spatial 
distribution of neutral hydrogen in  $\Lambda$CDM   models 
are also able to reproduce the large observed transverse correlation length  
of the Lyman-$\alpha$  forest in the absorption spectra of QSO pairs  
(Bi 1993; Miralda-Escud\'e et al. 1996;  Charlton et al. 1997; 
Viel et al. 2002; Rollinde et al. 2003; Rauch et  al. 2005).  

\parn

As pointed out by several authors  a comparison of 
the transverse correlation to the correlation observed along the 
line of sight, can be used to carry out a variant of the Alcock \&
Paczy\'nski (1979)  test to put constraints on the geometry of the 
Universe  (Hui, Stebbins \& Burleset 1999; Mc Donald \&  Miralda-Escud\'e 1999; 
Mc Donald 2003). This provides strong motivation for an (accurate) 
measurement of the transverse correlation function. 

\parn
In a previous work we have used 5 pairs and a group of 10 quasars 
with separations in the range 1-10~arcmin to investigate whether the 
longitudinal and transverse  correlation functions were consistent with 
those  expected in $\Lambda$CDM models (Rollinde et al. 2003). 
We reported  a somewhat marginal detection  of a transverse  
correlation up to separations of 3 to 4 arcmin.  We have 
assembled here a significantly larger sample of 32 QSO pairs.  
The new sample consists of  26 pairs 
with separations in the range 0.6$-$4~arcmin (corresponding to $\simeq$ 0.2 to 
1.4 $h^{-1}$~Mpc proper at z=2.1 for $\Omega_{\rm m}=0.3$, $\Ol=0.7$) 
and 6 pairs with separations in the range 5$-$10~arcmin.

Details of the observations and simulations are given in Sections 2 and 3, 
respectively. 
In  Section 4 we define and discuss our measurements of the 
observed longitudinal and transverse flux 
correlation functions. Section 5 compares the observed and 
simulated correlation functions. We  investigate the  
transverse correlation of  C~{\sc iv} absorption systems 
in Section 6. Our conclusions are given in Section 7.
Comments on individual lines of sight are given in Appendix A. 
Metal-line lists and QSO spectra are given in Appendix B and C
available in the electronic version of the paper. 

\section{Observations}

 \begin{figure*}
\vskip 1cm
 \unitlength=1cm
\vbox to220mm{\vfil{\psfig{height=22.cm,width=18.5cm,angle=-90,figure=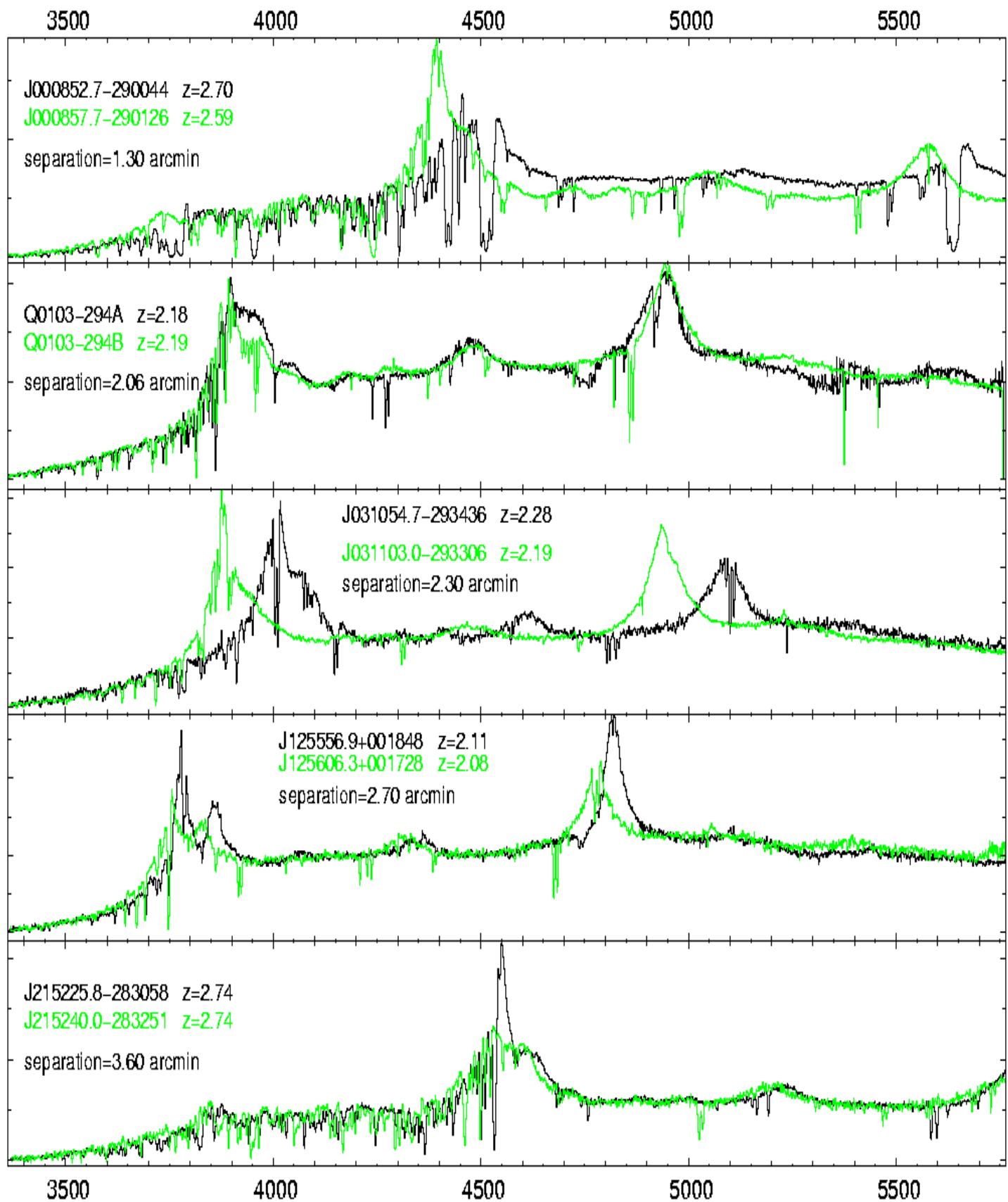 }}
 \caption{Typical observed spectra of QSO pairs in order of increasing right ascension.
 QSO names, emission redshifts and angular separations are given in the top-left corner 
of each sub-panel. Other spectra are presented in Appendix C published in the electronic version
of the paper.
 }
\label{f:pairs1}
\vfil }
 \end{figure*}


The first release of the 2dF quasar survey has significantly 
enlarged the number of known quasar pairs with arcmin 
separation (Outram, Hoyle \& Shanks  2001).  
We have selected pairs with the following criteria to enlarge the 
number of small separation pairs with respect to the sample of Rollinde et al. 
(2003):
(i) the separation of the two quasars should be in the range 1$-$4~arcmin 
where the correlation is expected and observed to be strong;
(ii) the quasars should be brighter than $m_{\rm V}$~=~20.30 to
keep observing time in reasonable limits;
(iii) the emission redshifts of the two quasars should be larger than $z$~$\sim$~2.1
to increase the wavelength range over which 
high S/N ratio can be obtained (FORS is not sufficiently sensitive  below 
3500~\AA);
(iv) the redshift difference should be smaller than $\Delta z$~$\sim$~0.5
(for most of them 0.3) to maximise the wavelength range over which 
correlation can be studied. 

There are 22 quasar pairs in the 2dF survey 
which meet our criteria of which we observed 20. We have 
observed two additional pairs not contained in 2dF : 
J~123510.5-010746$-$J~123511.0-010830  with a separation of 
0.74 arcmin and Q~1207-1057$-$Q~1206-1056 with a separation of 3.5 arcmin.
 The spectra were obtained with 
FORS1 and 2 mounted on the VLT-UT2 and UT3 telescopes of ESO  using the grism 
GR630B  and a 0.7~arcsec slit.  The spectra were reduced  using
standard procedures  available in  the context 
LONG of the  ESO data reduction package
MIDAS.  Master bias and flat-fields were produced using  day-time 
calibrations.  Bias subtraction  and flat-field  division  were 
performed  on science  and calibration images. A correction  for 
2D  distortion was applied.  The sky level  was evaluated
in  two windows on both  sides  of the  object offset  along the  slit
direction   and  subtracted  on  the  fly   during  the  optimal
extraction of the object.  The spectra  were  then wavelength  
calibrated  over the  range 3400~$<$~$\lambda$~$<$~6000~\AA.    
The  final   pixel  size   is  1.18~\AA~ corresponding  to a  resolution 
of  $R$~=~1400 or  FWHM~=~220~km~s$^{-1}$ at
3800~\AA.  The exposure times have  been adjusted in order to obtain 
a typical signal-to-noise ratio of $\sim$10 at 3500~\AA.  
The sharp decrease of the detector sensitivity below
4000~\AA~ prevents  scientific analysis below $\sim$3500~\AA. 
At $\lambda$~$\sim$~4500~\AA~ the S/N ratio is usually 
larger than 70.  
The final sample consists of 58 QSOs (44 QSOs new from this
program, 12 QSOs from Rollinde et al. 2003 plus the pair UM680-UM681
presented in D'Odorico et al. 2002). The total number of pairs
included in our analysis of the transverse correlation function 
is 32 somewhat larger than half the total number of QSOs due 
to the additional  pairings of Q~0103$-$294A\&B, Q~0102$-$2931 
and Q~0102$-$293 which form a group (alternative names are,
respectively,  J010534.7-290917, J010538.3-291106, J010518.0-291510
and J010502.8-290618).
One further pair in the sample (J~123510.5-010746$-$J~123511.0-010830) 
was at the end not included in our analysis of the transverse
correlation function as the redshift overlap 
of the Lyman-$\alpha$ forest is too small to contribute in  
a statistically significant way.

\Tab{pairs} gives a summary of the sample including 
emission redshifts, angular separation of 
the pairs on the sky, the mean S/N ratio in the wavelength range
of interest.  Emission redshifts were determined by fitting a 
Gaussian function to the \cIV emission line when present in the
spectrum or to the Lyman-$\alpha$ emission line otherwise.
Typical spectra are shown in Fig~ \ref{f:pairs1}. 
Other spectra are presented in Appendix C published in the electronic version
of the paper.
\par\noindent
The spectra have been normalized using a spline fit to the continuum. 
This operation is important as it can affect the estimate of the 
mean flux which is in turn critical for the flux correlation function estimate.
In our sample most of the Lyman-$\alpha$ forest common to two QSOs lies 
at $z$~$<$~2.3. At this redshift the density of absorption features is 
moderate and there are numerous spectral regions with no absorption. 
This is why continuum fitting is reliable. For the same reason,
the identification of metal absorption systems in the 
Lyman-$\alpha$ forest region of the spectrum is  
unproblematic.
We have checked that the mean absorption of the spectra in our sample 
is consistent with that  measured from the data of the 
VLT Large Program  (LP) 'The Cosmic Evolution of the IGM'
(Aracil et al. 2004, 2006 and below). 
\par\noindent
To calculate the longitudinal correlation function we also use the data 
from the  LP 'The Cosmic Evolution of the IGM' 
which has produced a sample of absorption spectra of homogeneous quality suitable
for  studying the Lyman-$\alpha$ forest in the redshift range 1.7$-$4.5. 
The spectra of the LP have been taken with VLT-UVES and have 
high resolution ($R$~$\sim$~45000), high signal to noise ratio
(30 and 60 per pixel at respectively 3500 and 6000~\AA) 
and cover the  wavelength ranges 3100--5400 and 5450--9000~\AA.
Details of the data reduction and normalization of the spectra 
are given in Aracil et al. (2004, 2006). 
\section{Numerical simulations}
\label{s:numericalsimulations}
In this paper we use two numerical simulations to estimate the  
errors and the effect of numerical resolution, redshift distortion 
and the thermal state of the gas on the correlation functions:  
a large size  dark-matter only simulation and a smaller size 
full hydrodynamical simulation.  For both  simulations 
we assume parameters consistent with the fiducial concordance cosmological 
model  $\Omega_\Lambda=0.7$, $\Omega_{\rm m}=0.3$. Hubble constants
are  $h=0.65$ and $h=0.7$ and normalization of the fluctuation
amplitude of the matter power spectrum are $\sigma_8=0.93$ and 
$\sigma_8=1.0$ for the  dark-matter only and the full hydrodynamical
simulation,  respectively. The assumed baryon density in the
hydro-dynamical simulations is  $\Omega_{\rm b}$~=~0.04.  
The simulations were performed on the computers of the Institut du
Developpement et des Ressources en Informatique Scientifique (IDRIS)
in Orsay.  

The hydro-dynamical simulation of 40 Mpc box-size and
dark-matter only simulation of 100 \hMpc~ box-size are used to catch both the
statistical aspects and the effect of gas physics. 
In the hydro-simulation some weak bias on the correlation functions 
due to the box-size is expected at separation larger than 4 Mpc. 
However, larger hydrodynamical simulations that still resolve 
the Jeans length at least marginally are currently not feasible.

The dark-matter only simulation was  performed with the
Particle~-~Mesh (PM) code described in Pichon et
al. (2001) and was also used in Rollinde et al. (2003). The simulation 
has 16 million particles and a box-size of 100 \hMpc.\footnote
{Note that the mean wavelength range corresponding
to the Lyman-$\alpha$ forest in the observed FORS spectra, $\sim$250~\AA,
corresponds to approximately twice the  box-size of the simulation.}  
The large box-size ensures a sufficient statistical sampling on large
scales where thermal  effects and pressure effects which are modelled
approximately in the dark-matter only simulation are less important. 
Initial conditions were set up using a standard CDM transfer function 
(Efstathiou, Bond \& White 1992).  To construct mock Lyman-$\alpha$
spectra from the simulated data, we proceed as in Rollinde et al. (2001, 2003) applying
simple semi-analytical prescriptions taking into account thermal broadening
and redshift distortion. The density and velocity fields are
interpolated on a 256$^3$ grid. We use adaptive smoothing similar to
SPH smoothing but with a Gaussian window truncated at $3 \sigma$, 
as explained and tested in Pichon et al. (2001). This eliminates discreteness
effects while keeping the best spatial resolution possible. 
The pixel size of the dark-matter only simulation is 0.4 \hMpc. 
This corresponds to 0.47 \AA,  a factor 2.5 smaller than the pixel size 
of the  FORS spectra. We have convolved the mock spectra from the numerical
simulations with a Gaussian filter to match the spectral resolution 
of the observed spectra. 

The hydrodynamical simulation is better suited  for investigating the 
effects of thermal broadening  and redshift distortions which are more relevant
on small scales. This simulation  has 512$^3$ dark
matter particles and follows the gas dynamics on a fixed cubic grid 
with 512$^3$ cells. The simulation has a box-size of 40 Mpc; the mesh 
size is $\sim$ 80 \hkpc~which corresponds to a pixel size of 
0.07 \AA~ at a redshift $z=2$.
This means that the Jeans length of the warm photo-ionized IGM is marginally
resolved.   
We have used simulations with box-size of 20 and 10 Mpc to check that the
40 Mpc simulation used here is not significantly affected by the fact
the Jeans mass is only marginally resolved. The 40 Mpc size therefore
offers the best compromise between box-size 
and resolution and the statistical  properties of the 
absorption spectra studied here are sufficiently converged.
The dark matter distribution is modelled with 
the same PM code as the dark-matter only simulation.  
The hydrodynamical part of the code is the same as in Chi\`eze, Alimi \& Teyssier (1998) 
and Teyssier, Chi\`eze \& Alimi (1998). The adiabatic hydrodynamic
step  is solved using directional splitting  and a staggered mesh.
Shock waves are approximated  with the pseudoviscosity method
(Von Neuman \& Richmyer 1950). An additional  dissipative step 
models the physical processes relevant for the description of gas dynamics
in a photoionized intergalactic medium, as described in the Appendix B of
Theuns et al. (1998), except that we use the heating and photoionization rates
of Dav\'e et al. (1999) which were derived from measurements by Haardt \& Madau (1996).
The Intergalactic Medium is highly ionized at the relevant redshifts 
and the dynamical evolution of the gas in the simulation depends therefore
only very weakly on the amplitude of the ionizing flux characterized
by its value at the Lyman limit: $J_{21}$.  
We have run the simulation with the same ionizing flux as
adopted in Dav\'e et al. (1999). However, when we are producing 
mock spectra, we compute the equilibrium ionic abundances 
in a post-processing step for a rescaled ionizing flux such as to
match  the observed flux distribution (see section \ref{s:numsimtest}).
This procedure does not affect the density and temperature distribution of 
the gas, as specified in Theuns et al. (1998).
%
\par\noindent
\section{The observed longitudinal and transverse correlation  functions}
\label{s:lya}
\subsection{Calculating correlation functions}
We define the  flux correlation  function as in Rollinde et al. (2003): 
\begin{equation} \xi_{f}(\theta,\Delta v)=\left\langle
({\cal     F}(\theta,\lambda+\Delta\lambda)-\langle{\cal    F}\rangle)({\cal
F}(0,\lambda)-\langle{\cal                  F}\rangle)\right\rangle_{\lambda}
\,,\EQN{correlation} 
\end{equation} 
where ${\cal  F}$ is the  normalized flux along  two \loss\ with
separation $\theta$  at a    mean     redshift      $z$;
$\Delta\lambda=\lambda_0(1+z)\times\Delta   v/c$, with $\lambda_0=1215.67$ \AA~
the hydrogen Lyman-$\alpha$ rest-wavelength, and   $c$ denotes the speed of light.
The velocity  distance corresponding to the angular separation $\theta$ 
can be written as $\Delta v_{\perp}=c\,f(z)\,\theta$,  where 
$f(z) = c^{-1} H(z) D_A (z)$,  $H(z)$ is the Hubble constant at $z$,
and $D_A(z)$  is the angular diameter distance (see Mc Donald 2003).
For $\theta=0$  equation (1) gives the  longitudinal correlation function. 
In the following we will use $H_{\rm 0}$~=~70~km/s/Mpc, $\Omega_{\rm m}$~=~0.3, 
$\Omega_{\Lambda}$~=~0.7 to relate scales in redshift space with
angular scales.  With these parameters $\Delta v$~=~100~km~s$^{-1}$ 
corresponds to $\sim$1~arcmin  at $z$~=~2. 
\par\noindent
We have excluded the wavelength range less than 1000~\kms\ redward 
of the Lyman-$\beta$ emission line when calculating the observed 
correlation functions  to avoid  contamination from the Lyman-$\beta$
forest. Likewise spectral regions less than 3000~\kms\ blueward of the  
Lyman-$\alpha$ emission line have been excluded to avoid the proximity
effect (see Rollinde et al. 2005).  Damped absorption systems and
metal lines which we were able to identify have been removed. 
The properties of identified damped absorption systems are listed in \Tab{DLA} 
and the properties of the metal lines are listed in the Appendix B published
in the electronic version of the paper.
\par\noindent
\subsection{The observed longitudinal correlation function}
\label{s:longcorr}
The observed correlation functions depend strongly on the spectral
resolution of the absorption spectra unless the width of all 
spectral features is fully resolved (e.g Becker, Sargent \& Rauch 
2004). 

\begin{figure}
\unitlength=1cm
\begin{picture}(9,8)
\centerline{\psfig{width=7.5cm,figure=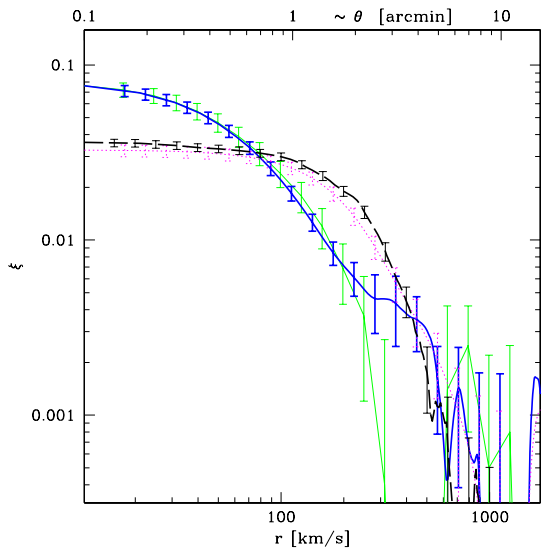}}
\end{picture}
\caption{The observed longitudinal correlation functions 
derived from low resolution FORS spectra (thick dashed curve),
high-resolution VLT-UVES (thick solid curve) and high resolution 
KECK-HIRES spectra (Mc Donald et al. 2000, thin solid curve). 
The thin  dotted curve shows the correlation function from the VLT-UVES spectra 
after convolution of the spectra with a Gaussian filter of 
FWHM~=~220~km~s$^{-1}$ to match the resolution of the FORS spectra. 
Errors for the FORS and UVES spectra have been estimated from the observations
as described in the text.} 
\label{f:longcorr}           
\end{figure}

We first consider the longitudinal correlation function 
obtained from the FORS data.  The thick dashed line  in Fig.~\ref{f:longcorr}
shows the average 
of the longitudinal correlation functions measured from the  58 FORS 
spectra in the sample. Errors are calculated from the variance
of the measurements (see Section \ref{s:errors}).

The longitudinal correlation  function measured from the high resolution 
spectra obtained in the course of the UVES-VLT LP is shown
as a thick solid curve in Fig.~\ref{f:longcorr}.  The high-resolution
spectra have a mean redshift of 2.39. As expected, the
correlation function of the high-resolution spectra extends to higher
values at small velocity separation.  We also show the longitudinal 
correlation function obtained by Mc Donald et al. (2000) from a  
smaller sample of  eight high-resolution spectra with a  mean redshift of 
$z$~=~2.41  obtained with Keck-HIRES as the thin solid curve. There is 
excellent agreement for the two samples of high-resolution spectra at
velocity separations $\le$~300~km~s$^{-1}$.
At velocity separations of $\sim$~400~km~s$^{-1}$, the longitudinal
correlation function obtained from the UVES spectra appears  to be 
larger than that obtained from  the HIRES spectra. 
Cosmic variance or artefacts due to 
continuum fitting errors are two plausible explanations.
Note, however, that the errors are already rather large at these
separations and that the difference is probably not statistically
significant.   
   
The thin dotted curve shows the longitudinal correlation from 
the high-resolution VLT-UVES spectra after convolution 
with a Gaussian  filter of FWHM~=~220~km~s$^{-1}$  
to take into account the difference in resolution of the UVES and 
FORS spectra. It agrees very well with the FORS
correlation function up to a small systematic offset. 

\subsection{The observed transverse  correlation function}
\label{s:transcorr}
\begin{figure}
\unitlength=1cm
\begin{picture}(9,8)
\centerline{\psfig{width=7.5cm,figure=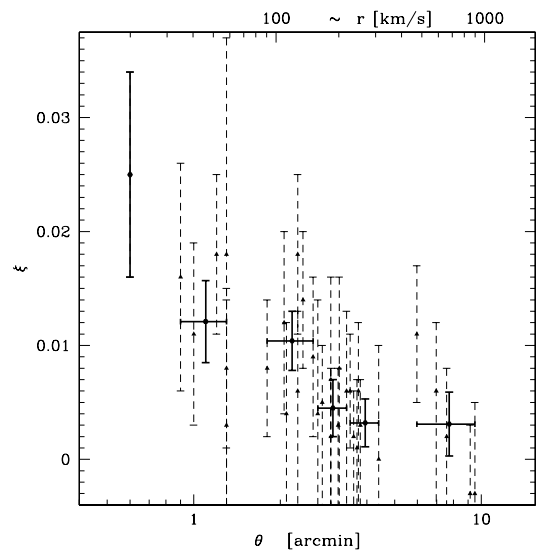}}
\end{picture}
\caption{ The observed transverse correlation coefficient  for individual pairs 
(black triangles, dashed error bars; see Table 1) and a binned estimate of the
transverse correlation function (solid error bars). 
The error bars for the individual measurements are estimated using  
the dark-matter only simulation as explained in Section \ref{s:errors}
}
\label{f:comparison1}         
\end{figure}

\parn 

The correlation function is calculated using the 32 pairs presented in 
Table~1. We use only spectral regions where the S/N ratio in both spectra 
is larger than 8. 
For this reason the pair J~123510.5-010746$-$J~123511.0-010830 
is not included in the analysis as the redshift overlap 
of the two Lyman-$\alpha$ forests is too small to contribute in  
a statistically significant way. The wavelength range 
($\lambda_{\rm max}-\lambda_{\rm min}$) used to compute the 
transverse correlations is given in \Tab{pairs} together with  the corresponding 
number of pixels. The mean flux is taken over each individual line of sight.

In Fig.~\ref{f:comparison1} we show the observed transverse correlation
function. The measurement for each quasar
pair, $\chi(\theta_{\rm i}) \equiv \xi_{f}(\theta_{\rm i},\Delta v = 0)$, 
is shown  as a small solid triangle at the angular separation of the
pair, $\theta_{\rm i}$. 
The points with the solid error bars are a binned estimate of the transverse 
correlation function for which we have weighted the individual measurements with  the 
inverse of their  errors (see Section \ref{s:errors} for the computation 
of errors). Note that the first bin at the smallest separation 
contains only one measurement. 

The transverse correlation is clearly detected on scales $<4$~arcmin.  
If we merge the two bins between 3 and 4 arcmin, the correlation is detected
at about the 3$\sigma$ level.

Note that correlation coefficients derived from this work for some of the pairs
observed by Rollinde et al. (2003) differ somewhat from what is given in 
their paper (see Table 1). This is due to slight differences in 
the reduction of the data and the determination  of the continuum.

\subsection{Estimation of errors}
\label{s:errors}

The measurements of the longitudinal correlation reported in Fig.~\ref{f:longcorr} 
are the average of the correlation function of the individual
spectra $\xi_i(\Delta v)$ (58 spectra in the case of FORS and 20 in 
the case of UVES), 
\begin{equation}
{\bar \xi}(\Delta v)=\frac{1}{N} \sum_{i=1}^N \xi_i(\Delta v)\, . 
\label{e:simpav}
\end{equation}
\parn
The error of the mean correlation function is then computed as, 
$(\Delta \xi)^2 \simeq  1/N(N-1) \sum_{i=1}^N [\xi_i-{\bar \xi}]^2.$
For the transverse correlation there is only one measurement 
(one pair) at each separation.  We therefore use the dark matter-only  
simulation to estimate the statistical errors.
We choose samples of random sightlines along one axis of the box 
carefully reproducing individual pair separations, wavelength coverage, 
resolution and the noise of the spectra in the observed sample.  
The length of the observed spectra is always larger than 100 \hMpc,
and we have concatenated  several lines of sight through the
simulation box in order to reproduce one observed spectrum.
Note that we use only one output of the simulation at $z=2$
and  did not try to account for the moderate redshift 
evolution in our sample. We extract 10000 different realizations 
from the simulation.  We then fit a Gaussian to the distribution of 
the values of the correlation function at each separation 
${\hat \xi}_j$  and use the rms of the distribution 
as estimate for the error of the transverse correlation function 
as reported in \Tab{pairs} and Figure \ref{f:comparison1}.

Note that the errors quoted here on the flux correlation functions are
only indicative because they are strongly correlated (see McDonald et al. 2000).
\section{Comparison of observed and simulated correlation functions} 

\subsection{Numerical simulations as a testbed for systematic uncertainties}
\label{s:numsimtest}

The longitudinal and transverse correlation functions reflect the
clustering of the underlying matter distribution in real space 
albeit in a somewhat indirect way. The correlation functions
calculated from mock spectra  produced from numerical simulations 
are an excellent tool to test the effects of resolution, redshift space 
distortion, thermal broadening and non-linear evolution of the gravitational clustering. 
We use here the 512$^3$ cell full hydro-simulation described in \Sec{numericalsimulations}.
When thermal broadening and redshift-distortion are
taken into account, they are computed from the temperature and
velocity fields of the simulation as described in Theuns et al. (1998).
We produce spectra for all lines of sight along one axis of the
simulation box separated by one cell. This corresponds to
512$^2$ sightlines with a length of 512 pixels each.
Our estimate of the longitudinal correlation function from the
simulations is obtained by averaging over these 512$^2$ individual 
realisations. The transverse correlation function is computed
at 20 log-spaced values of $\theta$.  We average over  pairs of lines
of sight for each value of $\theta$ in the following way. For each
of the $512^2$ pixels in  the y-z plane, we take the sightline parallel
to the x-axis as the first spectrum of a pair. We then use  two
parallel lines of sight separated by a distance $\theta$,
in the y-direction and z-direction, respectively,  to compute the
second spectrum to obtain  two pairs of spectra. Our estimate of the transverse
correlation is the  average of the resulting $2\times 512^2$ pairs. 
\parn The normalization of the correlation function 
depends sensitively on the mean flux which in turn depends on the 
amplitude of the ionizing flux. Therefore, the mock spectra 
were calculated with a rescaled ionizing flux such that the  
probability distribution function (PDF)
of the flux distribution matches that of our observed spectra 
at the same redshift for $\tau$~=~1. 
We proceed iteratively starting with an arbitrary ionizing
flux and adjusting this flux step by step till the fit is obtained.
We found that this procedure is similar although more robust than 
the conventional procedure to match the mean flux in the Lyman-$\alpha$ forest. 
Indeed, the fit of the PDF minimizes the role played by overdensities and
the effect of cosmic variance.

\begin{figure}
\unitlength=1cm
\begin{picture}(9,10.5)
\put(0,0){\psfig{width=7.5cm,figure=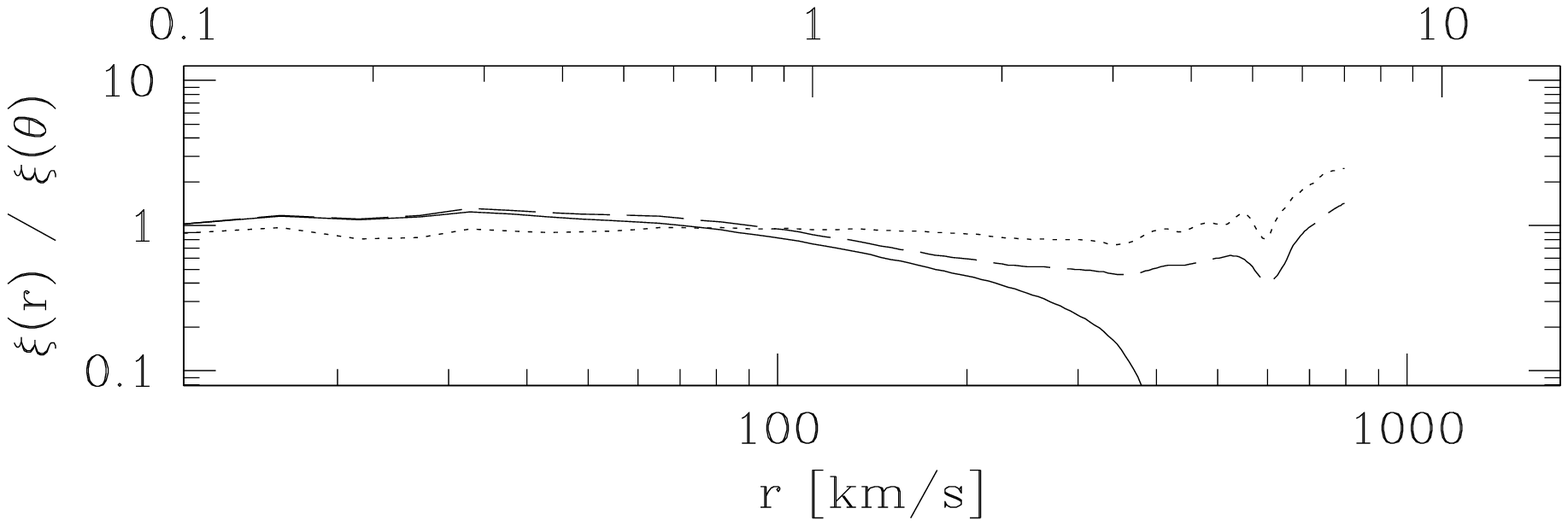}}
\put(0,2.5){\psfig{width=7.5cm,figure=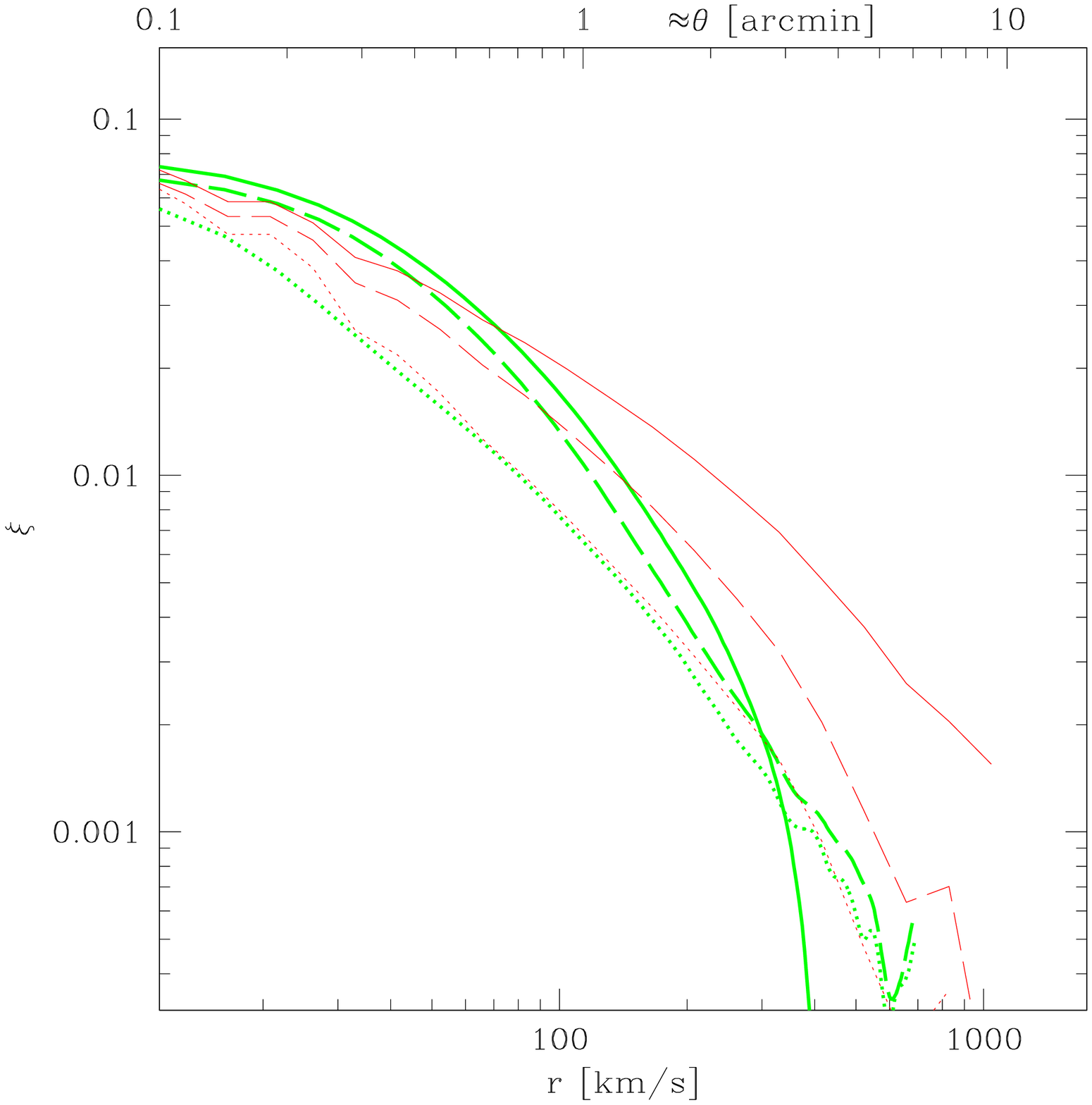}}
\end{picture}
\caption{The longitudinal and transverse correlation functions 
versus velocity separation (lower x-axis) and  angular separation 
(upper x-axis) for the full hydro-simulation at $z$~=~2. 
{\sl Upper panel}: the longitudinal correlation of the 
gas density (thick dotted curve) is nearly identical to the transverse 
correlation at the same redshift (thin dotted curve) as expected. The 
correlation functions differ when thermal broadening  (longitudinal:
thick dashed curve, 
transverse: thin dashed curve) and peculiar velocities 
(longitudinal: thick solid curve, transverse: thin solid curve) are 
taken into account. 
{\sl Lower panel}: Ratio of the longitudinal and transverse
correlation functions. Linestyles are the same as in the upper panel. }
\label{f:sim_corrf}         
\end{figure}

The thick and thin dotted curves in Fig.~\ref{f:sim_corrf} show,
respectively, the
longitudinal and transverse correlation functions in real space
as calculated from mock spectra produced from the full
hydro-simulation.  We have again used our fiducial cosmological 
parameters to relate velocity and angular separation.  As expected the 
two correlation functions  are almost identical. The dashed curves show 
the same comparison with the effect of thermal broadening included. There are
now significant differences. At small scales the longitudinal
correlation function exceeds the transverse correlation function while
at large scales the opposite is true. The solid curves show the
effect of including peculiar velocities. The corresponding redshift
distortion further enhances the differences between longitudinal and
transverse correlation functions. The scale dependence of the
difference is similar to that due to thermal broadening but the
differences are significantly larger especially at scales larger 
than  200~km~s$^{-1}$  or 2 arcmin. 


A proper quantitative
understanding of these effects with the help of numerical simulations 
will be essential  for attempts to use  the comparison of observed 
longitudinal and transverse  correlation functions to 
measure cosmological parameters. This would need a full set
of simulations spanning the whole range of parameters and is 
therefore beyond the scope of this paper.

\subsection{Observed {\it vs} simulated correlation functions}
\label{s:simfunc}
\parn 

\begin{figure}
\unitlength=1cm
\begin{picture}(9,8)
\centerline{\psfig{width=7.5cm,figure=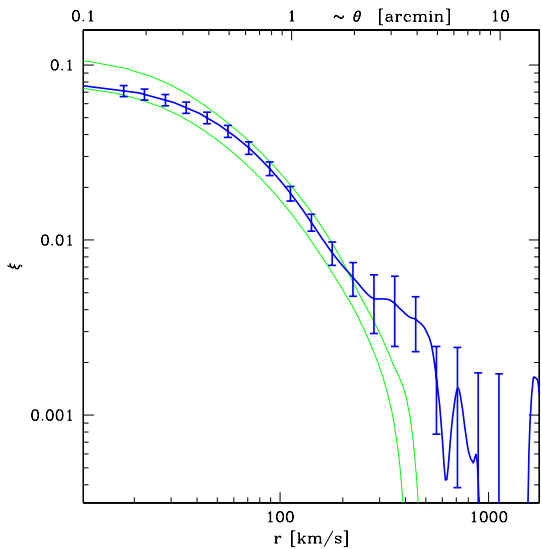}}
\end{picture}
\caption{The observed longitudinal correlation function from the high 
resolution UVES spectra (thick solid curve) compared to the 
longitudinal correlation function as measured in the full 
hydro-dynamical simulation at $z=2$ (lower thin curve)
and $z=3$ (upper thin curve), respectively. }
\label{f:convolution}
\end{figure}

The ability of $\Lambda$CDM  models to reproduce the longitudinal
correlation function of the Lyman-$\alpha$ forest has been demonstrated
by many authors (e.g. Croft et al. 2002, Viel et al. 2002,
Rollinde et al. 2003) and, as we will see below, the same is true 
for our simulations. The thin solid curves in Fig.~\ref{f:convolution}  
show the mean longitudinal correlation function obtained for mock spectra
produced from the hydrodynamical simulation at $z=2$ (lower curve) 
and $z=3$ (upper curve).  The curves nicely bracket the observed 
correlation function obtained from the high-resolution data 
with a median redshift of $z=2.39$. Note again the slight excess of the 
longitudinal correlation function of the UVES data at large scales which is, 
however,  probably not statistically significant. 

\begin{figure}
\unitlength=1cm
\begin{picture}(9,8)
\centerline{\psfig{width=7.5cm,figure=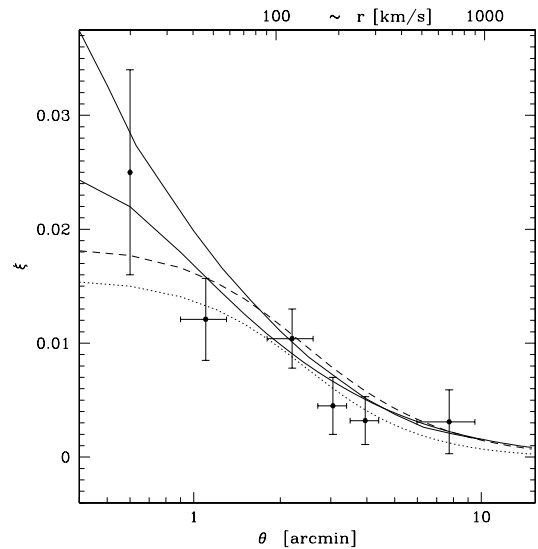}}
\end{picture}
\caption{
The binned estimate of the observed transverse correlation function
(solid error bars) is shown together with the estimate 
of  the transverse correlation function from 
the full hydro-simulation (thick solid curve) and linear
predictions for $\Omega_{\rm m}$~=~0.1, 0.3 and 1 (thin solid, dashed 
and dotted lines respectively, assuming a flat universe : 
$\Omega_{\Lambda}+\Omega_{m}=1$; see text). The normalization of the flux 
in the simulation  
is fixed in order to reproduce  the observed flux PDF (Section \ref{s:simfunc}). 
The linear theory predictions are  normalized to reproduce the 
longitudinal correlation function at large scales. 
}
\label{f:comparison2}         
\end{figure}

\par\noindent

The  transverse correlation contains precious direct information on 
the physical size/coherence-length of the absorbing
structures as it is less affected by redshift space distortions
than the longitudinal correlation function. Furthermore, 
a comparison of the longitudinal and transverse correlation 
functions can -- at least  in principle -- strongly constrain
cosmological parameters in particular $\Omega_{\Lambda}$.

The thick solid curve in Figure \ref{f:comparison2} shows our estimate of
the transverse correlation function from the  full hydro-dynamical 
simulation at $z=2$. It agrees well with our measurement of 
the observed transverse correlation function (mean redshift $\sim$2.1) 
which is shown as the solid dots with error bars. 
The thin dashed curve shows the prediction of linear theory (Kaiser 1987; 
Mc Donald \& Miralda-Escud\'e 1999) for  the cosmological parameters 
assumed for the hydro-simulation. 
The thin solid and dotted curves show the  prediction of linear
theory for $\Omega_{\rm m} =0.1$ and  $\Omega_{\rm m} =1.0$, respectively 
(assuming a flat universe : $\Omega_{\Lambda}+\Omega_{m}=1$ and 
adjusting other parameters to fit the data).  
The linear theory predictions
are normalized so that the longitudinal correlation
function is best fitted for $\Delta v> 200$ km\,s$^{-1}$.
As expected, the linear predictions agree reasonably  well  with 
the numerical simulation at large scales but underpredicts the 
correlation function substantially at small scales.  
The non-linear effects of gravitational clustering 
are clearly visible in the observed transverse correlation function. 

Despite the larger sample (about three times more pairs at 
$\theta<3$ arcmin than  in Rollinde et al. 2003) and the
correspondingly smaller errors, we cannot yet distinguish between different
values of $\Omega_{\rm m}$. This confirms the predictions by Rollinde
et al. (2003) and  Mc Donald (2003) that significant constraints 
on $\Omega_{\Lambda}$ require a larger number of
pairs. Using the (cross) power spectrum instead of the transverse and longitudinal 
correlation functions, Mc Donald (2003) estimates that  
of the order of 13($\theta$/1')$^2$ quasar pairs on scales  up  to 10~arcmin
are necessary to perform the test. In addition,
performing the Alcock \& Paczy\'nski test using the 
correlation functions at small scales ($\le$~3~arcmin or  300
km~s$^{-1}$ at $z$~=~2)  will require the use of a large suite 
of full hydro-dynamical simulations.

\parn 
%
\begin{table}
 \begin{center}
 \begin{tiny}
\begin{tabular}{|p{1.4cm} p{0.3cm} p{0.3cm} | p{0.2cm} p{1.cm} p{0.6cm} |  p{0.5cm} p{0.5cm} |}
 \hline
  QSO name& $z_{\rm em}$&$\Delta\theta$ & SNR & $\lambda$ & Number & $\xi(0,\theta)$&$\sigma$\\
   & & (arcmin) & & (\AA) &of pixels & & \\
 \hline
 \hline
 $*$Q2139-4504B        &   3.255   &    0.600   &  24.4  & 4381- 4878  & 420  & 0.025 & 0.009 \\ 
 ~~~Q2139-4504A        &   3.055   &            &  15.2  &             &      &       &  \\ 
\hline
 ~J~123510.5-010746        &   2.785   &    0.740   &  51.4  & 3897- 3904  & 7  & - & - \\ 
 ~J~123511.0-010830        &   2.235   &            &  26.2  &             &      &       &  \\ 
\hline 
 ~J~232800.7-271655  &   2.378   &    0.900   &  14.3  & 3756- 4038  & 240  & 0.016 & 0.010 \\ 
 ~J~232804.4-271713  &   2.357   &            &  14.3  &             &      &       & \\ 
\hline 
 ~UM680              &   2.144   &    1.000   &   50.0 & 3235- 3699&9641$^a$&0.011&0.008 \\ 
~UM681&2.122&&50.0&&&  \\ 
\hline 
~J~031036.4-305108&2.552&1.200&20.9& 3658- 4249&501&0.018&0.007 \\ 
~J~031041.0-305027&2.532&&25.4&&&  \\ 
\hline 
~J~135001.7-011703&2.657&1.300&15.5& 3765- 3821&48&0.018&0.019 \\ 
~J~135003.0-011819&2.177&&18.5&&&  \\ 
\hline 
~J~000852.7-290044&2.699&1.300&64.1& 3808- 4323&317&0.008&0.007 \\ 
~J~000857.7-290126&2.593&&34.2&&&  \\ 
\hline 
~J~214507.0-303046&2.532&1.300&37.5& 3636- 3869&177&0.003&0.011 \\ 
~J~214501.6-303121&2.216&&27.3&&&  \\ 
\hline 
~J~005852.4-272933&2.565&1.800&23.9& 3671- 4284&493&0.008&0.006 \\ 
~J~005859.1-273038&2.561&&25.5&&&  \\ 
\hline 
$*$Q0103-294B&2.190&2.063&25.5& 3602- 3827&191&0.012&0.008 \\ 
~~~Q0103-294A&2.182&&28.3&&&  \\ 
\hline 
$*$Q2129-4653B&2.222&2.100&22.4& 3603- 3856&215&0.004&0.008 \\ 
~~~Q2129-4653A&2.206&&16.8&&&  \\ 
\hline 
~J~031054.7-293436&2.281&2.300&18.3& 3601- 3833&197&0.006&0.007 \\ 
~J~031103.0-293306&2.187&&29.3&&&  \\ 
\hline 
~J~102827.1-013641&2.393&2.300&13.4& 3728- 3954&193&0.018&0.007 \\ 
~J~102832.6-013448&2.287&&17.0&&&  \\ 
\hline 
~J~111201.8-013018&2.549&2.400&48.7& 3654- 3954&246&0.014&0.006 \\ 
~J~111200.4-013242&2.292&&28.1&&&  \\ 
\hline 
$*$Q0236-2411&2.260&2.600&20.3& 3602- 3860&219&0.009&0.007 \\ 
~~~Q0236-2413&2.211&&19.4&&&  \\ 
\hline 
~J~125556.9+001848&2.108&2.700&19.9& 3602- 3708&90&0.004&0.010 \\ 
~J~125606.3+001728&2.083&&20.0&&&  \\ 
\hline 
~J~013734.2-303802&2.481&2.800&24.4& 3602- 4005&341&0.005&0.005 \\ 
~J~013734.2-304050&2.329&&29.2&&&  \\ 
\hline 
~J120725.9-024519&2.676&3.000&33.2& 3785- 3904&101&0.007&0.009 \\ 
~J120734.5-024725&2.245&&20.0&&&  \\ 
\hline 
~J~095810.9-002733&2.559&3.000&21.4& 3664- 4047&287&0.002&0.006 \\ 
~J~095800.2-002858&2.364&&14.2&&&  \\ 
\hline 
~J~223850.1-295612&2.448&3.180&32.0& 3602- 4062&390&0.003&0.005 \\ 
~J~223850.9-295301&2.377&&34.2&&&  \\ 
\hline 
~J~141124.6-022943&2.710&3.210&43.2& 3820- 3971&128&0.008&0.008 \\ 
~J~141117.3-023222&2.301&&35.3&&&  \\ 
\hline 
~J~023836.9-282310&2.565&3.400&31.7& 3677- 3899&175&0.006&0.007 \\ 
~J~023849.0-282101&2.242&&56.7&&&  \\ 
\hline 
~Q~1207-1057&2.450&3.500&33.8& 3601- 3975&318&0.006&0.005 \\ 
~Q~1206-1056&2.305&&24.7&&&  \\ 
\hline 
~J~215225.8-283058&2.741&3.600&38.0& 3851- 4494&545&0.002&0.004 \\ 
~J~215240.0-283251&2.736&&25.3&&&  \\ 
\hline 
~J~112116.1+003112&2.205&3.700&19.0& 3603- 3834&197&0.001&0.006 \\ 
~J~112108.2+003420&2.188&&22.8&&&  \\ 
\hline 
~J144245.7-023906&2.551&3.740&24.3& 3656- 4011&232&0.006&0.006 \\ 
~J144245.6-024251&2.334&&21.2&&&  \\ 
\hline 
~J~230318.4-290120&2.587&3.800&35.5& 3693- 4285&500&0.003&0.004 \\ 
~J~230301.6-290027&2.562&&26.9&&&  \\ 
\hline
 \end{tabular}
\end{tiny}
 \end{center}
\caption{Properties of the 33 QSO pairs: QSO names, emission redshifts, 
angular separation on the sky, mean S/N ratio over the wavelength range
of interest, wavelength range (in \AA) over which 
the correlation is calculated, corresponding number of pixels, value of 
the correlation function and errors (see Section 4.4).
(a)~UVES data (D'Odorico et al. 2002) with 0.04 \AA~ per pixel instead of 1.18~\AA~ for FORS data.
(*) QSO pair observed by Rollinde et al. (2003).
}
 \label{t:pairs}
 \end{table}

\begin{table}
 \begin{center}
 \begin{tiny}
\begin{tabular}{|p{1.4cm} p{0.3cm} p{0.3cm} | p{0.2cm} p{1.0cm} p{0.6cm} |  p{0.5cm} p{0.5cm} |}
 \hline
  QSO name& $z_{\rm em}$&$\theta$ & SNR & $\Delta\lambda$ & Number & $\xi(0,\theta)$&$\sigma$\\
   & & (arcmin) & &(\AA) &of pixels & & \\
\hline
 \hline
$*$FOCAP QSF:01&2.267&4.400&21.5& 3602- 3673&61&0.000&0.010 \\ 
~~~FOCAP QSF:04&2.054&&19.0&&&  \\ 
\hline 
$*$Q0102-2931&2.212&5.974&18.5& 3603- 3837&199&0.011&0.006 \\ 
~~~Q0103-294B&2.190&&25.5&&&  \\ 
\hline 
$*$Q0102-2931&2.212&6.977&18.5& 3602- 3827&191&0.006&0.006 \\ 
~~~Q0103-294A&2.182&&28.3&&&  \\ 
\hline 
$*$Q0102-293&2.441&7.585&24.8& 3602- 3827&191&0.002&0.006 \\ 
~~~Q0103-294A&2.182&&28.3&&&  \\ 
\hline 
$*$Q0102-293&2.441&9.152&24.8& 3603- 3837&199&-0.003&0.006 \\ 
~~~Q0103-294B&2.190&&25.5&&&  \\ 
\hline 
$*$Q0102-293&2.441&9.506&24.8& 3602- 3864&222&-0.003&0.008 \\ 
~~~Q0102-2931&2.212&&18.5&&&  \\ 
\hline 
 \end{tabular}
\end{tiny}
 \end{center}
\contcaption{ - Alternative names for FOCAP QSF:01 and FOCAP QSF:04 are 
J034105.1-445619 and J034126.2-445842. }
 \label{t:pairs2}
 \end{table}

\section{Metal absorption systems}

%
%
\par\noindent

\begin{figure}
 \unitlength=1cm
\begin{picture}(9,10)
\centerline{\psfig{height=10.5cm,width=8.5cm,figure=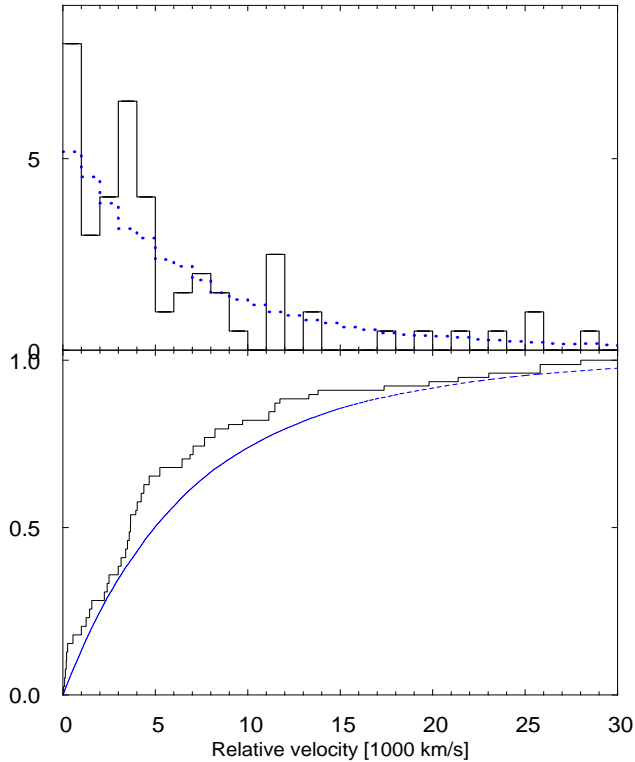}}
\end{picture}
 \caption{{\sl Upper panel:} Histogram of the velocity separations between 
two closest-neighbor C~{\sc iv} absorption systems in our sample of QSO pairs. The 
distribution expected from a randomly distributed population of C~{\sc iv} systems 
is shown as the dashed curve. {\sl Lower panel:} the  cumulative distributions
of the observed sample of C~{\sc  iv} systems  (thick curve) is
compared to the  cumulative distribution of a randomly distributed
population of C~{\sc iv} systems (dashed curve). There is a 18\%
chance probability (KS test)  that the distributions
differ that much if the two sample are drawn from the same population. 
}
 \label{f:Hist}
 \end{figure}
\par\noindent
\begin{figure}
 \unitlength=1cm
\begin{picture}(9,10)
\centerline{\psfig{height=10.5cm,width=8.5cm,figure=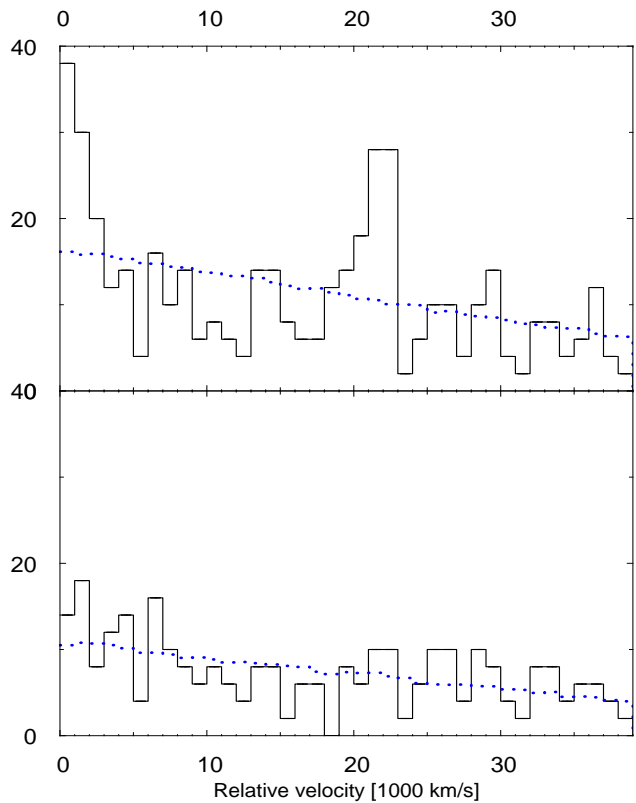 }}
\end{picture}
 \caption{Observed Longitudinal correlation function of C~{\sc iv} systems
computed along the 58 lines of sight. The distribution expected from a population 
of randomly distributed C~{\sc iv} systems is shown  as the dashed curve.
{\sl Upper panel}: the whole sample is used. There is an excess
on small scales and around 20000~km~s$^{-1}$. This is due 
to an excess of correlation along the lines of sight to Q~0103$-$294A,B; Q~0102$-$2931 
and Q~0102$-$293 located in the same field (group of quasars). {\sl Lower panel}: same  but
without the group.}
\label{f:corrC}
  \end{figure}

\par\noindent
\begin{figure}
 \unitlength=1cm   
\vbox {\vfil{\psfig{height=5cm,width=8.5cm,angle=0,figure=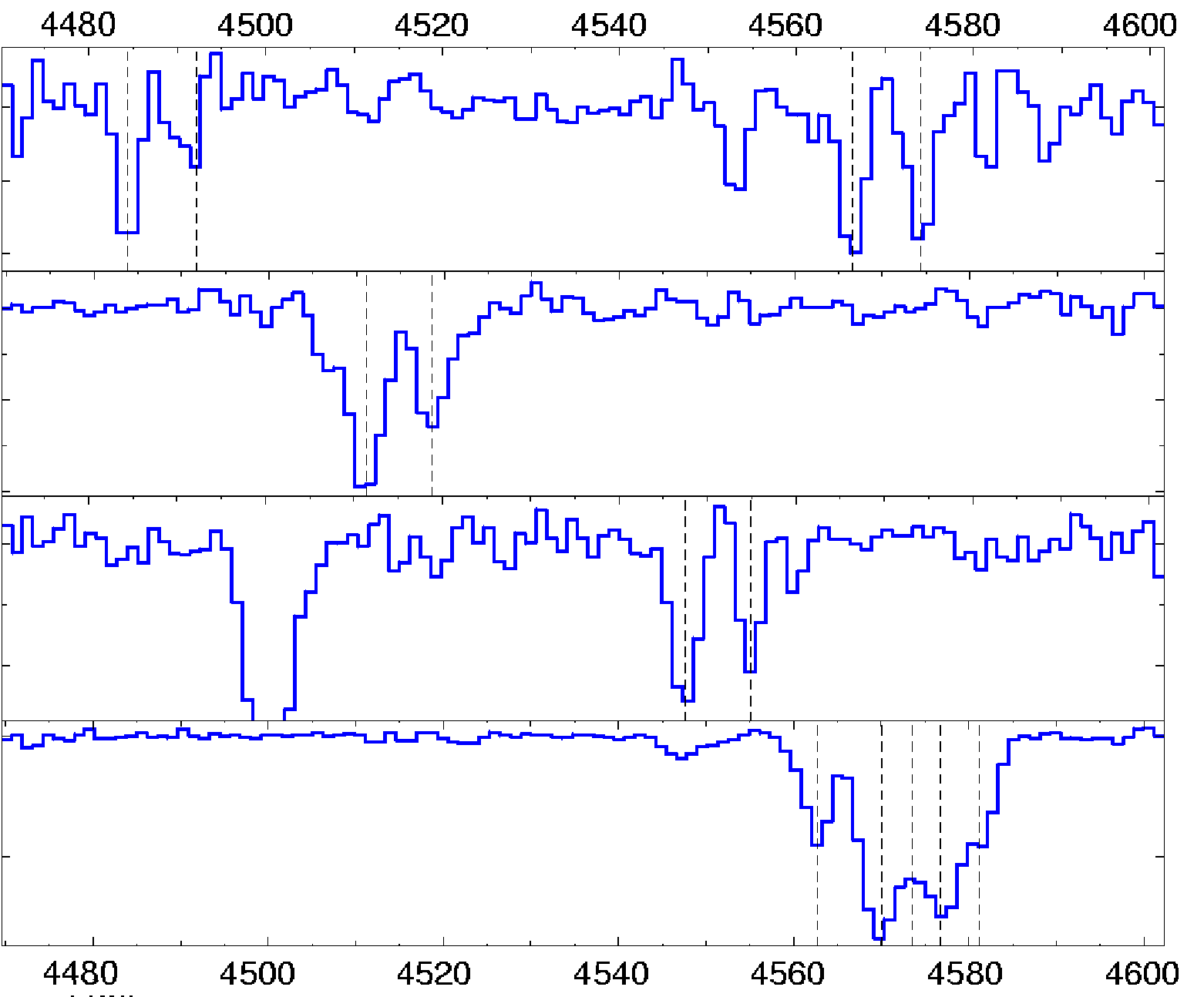}}
\vfil}
\vbox {\vfil{\psfig{height=5cm,width=8.5cm,angle=0,figure=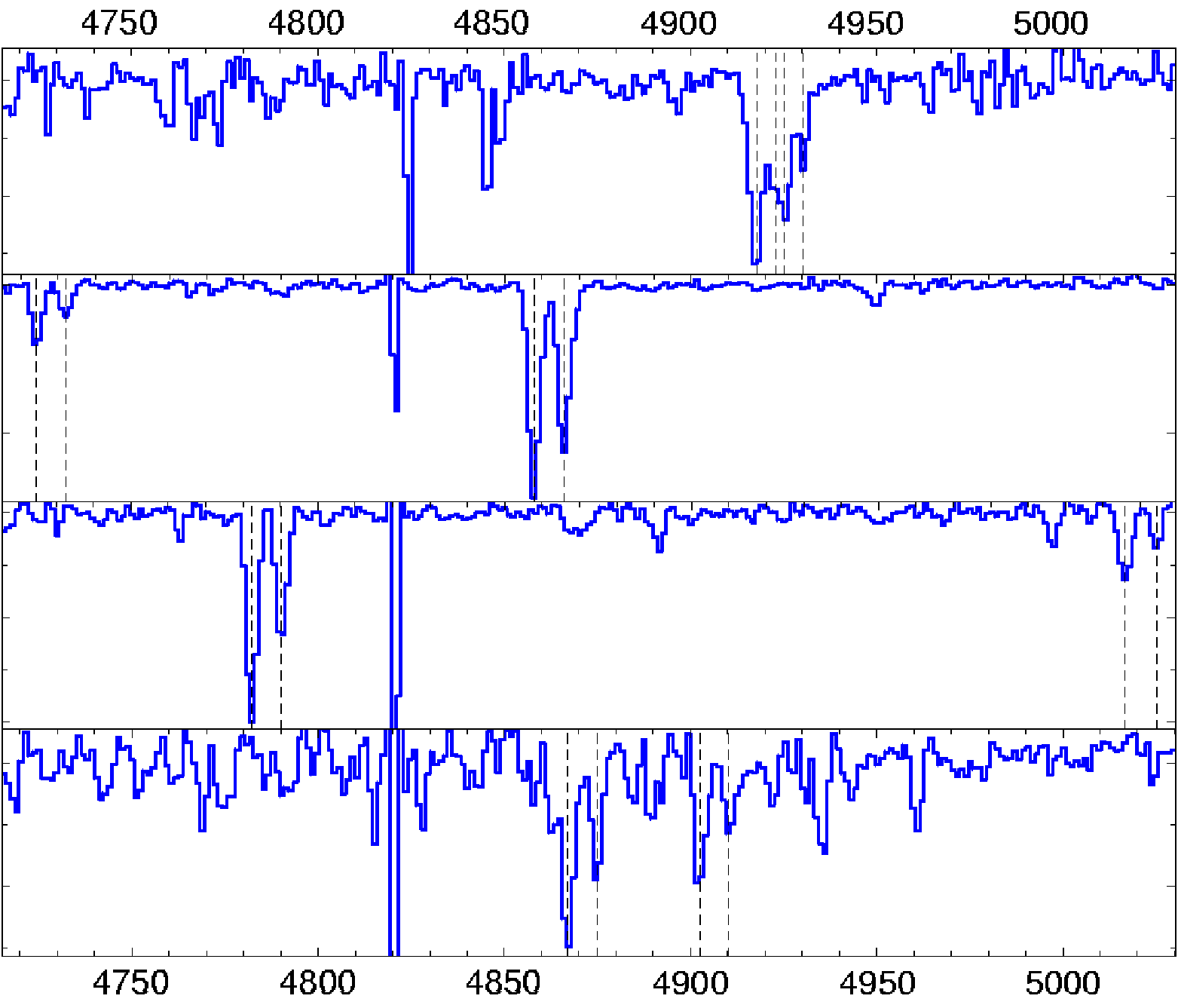}}
\vfil}
 \caption{Two sections of the spectra of Q~0103$-$294A\&B (separated by 1.3~arcmin) and
Q~0102$-$2931 and Q~0102$-$293  (separated by about 6.5~arcmin from
Q0103$-$294A,B , see Table~1) in that order from top to
bottom. The mean redshifts are 
1.93 and 2.14 in the upper and lower panels, respectively. The positions of 
C~{\sc iv} systems are indicated by dashed vertical lines.
There is an overdensity of C~{\sc iv} systems along these
lines of sight over the redshift range 1.5~$\leq$~$z$~$\leq$~2.2 and on a spatial
scale larger than 10~arcmin. 
}
\label{f:Spectra}
\end{figure}
\subsection{Identifying metal lines}

Our sample is also well suited to study the spatial distribution of
the gas responsible for the associated metal absorption in QSO
spectra. We have manually identified and fitted metal lines in all 
spectra.  The corresponding line lists  are compiled in 
Appendix B (published in the electronic version of the paper)
where we report the observed equivalent width, 
the pixel signal-to-noise ratio at the position  
of the absorption, the observed wavelength and corresponding redshift
for each of the identified metal lines. If applicable  
the upper limit on the equivalent width of a possible absorption
at the same position in the spectrum of the second quasar of the pair
is also given. 

\par\noindent
\subsection{The correlation of \cIV~ systems along adjacent lines of sight}
We consider only absorption lines with rest-frame equivalent width 
$W_{\rm r}$~$>$~0.1~\AA~ and redshift intervals common 
to both lines of sight of a QSO pair. We do not consider systems
where N~{\sc v} is detected as these systems are most probably 
associated with the quasar (see e.g. Petitjean, Rauch \& Carswell 1994).
We then select the lines that are located at more than 3000~\kms~ blueward of the 
QSO \cIV~emission line and  at more than 1000~\kms~ redward of the
Lyman-$\alpha$ emission line. We end up with a sample of 139 \cIV~
systems for a redshift path $\delta z = 38$, corresponding to a density of
3.7 systems per unit redshift. We apply the Nearest-Neighbor method, as described in Young 
et al. (2001) and Aracil et al. (2002) to the corresponding list of 
C~{\sc iv} systems. For each absorption line along one QSO line of sight, 
we search the adjacent QSO line of sight for the nearest (in velocity) 
absorption line and construct the histogram of the corresponding velocity 
differences (see Fig.~ \ref{f:Hist}). Our complete sample contains 25 and
39 associations with velocity separations smaller than 5000~and 
30000~km~s$^{-1}$, respectively. 
To estimate the possible excess of correlation with respect to randomly located 
absorption lines, we produced 10000 simulated line lists drawn  
from a population of randomly redshifted lines, taking  the same number of 
lines and the same wavelength ranges as in the observed 
spectra. The results of applying the same method to the simulated line
lists are given as dotted lines in Figure \ref{f:Hist}. 
\par\noindent
In the lower panel of Fig.~\ref{f:Hist} we compare the cumulative 
distributions of the velocity differences from our observed sample 
and from a randomly located population of \cIV~ absorbers. A KS test gives
a 18\% chance probability that the difference between the two distributions is
larger than what is observed if the two samples are drawn from the same population. 
There is a possible small excess of clustering of \cIV~ systems on scales smaller
than 5000~km~s$^{-1}$. 
This scale is larger than the typical correlation length,
of about 1000~km~s$^{-1}$, seen in the longitudinal correlation function of
C~{\sc iv} systems (Rauch et al. 1996; Pichon et al. 2003; Boksenberg, Sargent \& Rauch 2003;
Scannapieco et al. 2006). 
This is due to an excess of associations in the bin $\Delta v$~$\sim$~4000~km~s$^{-1}$.
The corresponding C~{\sc iv} associations are located in the peculiar field containing 
the quartet Q~0103$-$294A\&B, Q~0102$-$2931 and Q~0102$-$293 (all four quasars
are separated by less than 10~arcmin).
In fact, 9 out of the 25 associations with $\Delta v$~$<$~5000~km~s$^{-1}$ (see 
next Section) are found in front of these quasars. If we remove this quartet from the
sample, the KS probability is increased to 25\%.
To ascertain the overdensity in this field,
we show in Fig.~\ref{f:corrC} the longitudinal correlation function for the whole
sample (upper panel) and for the sample without the group (lower panel). 
There is indeed a strong excess in the correlation function of the whole sample
for $\Delta v$~$<$~3000~km~s$^{-1}$ and around 20000~km~s$^{-1}$ 
which disappears when the group is removed  from the sample.
\par\noindent
The pairs in the current sample have a mean separation larger than 2~arcmin
(see Table~1). The correlation at smaller separations can be expected
to be larger. Indeed, comparing the results of
$N$-body simulations to high spectral resolution
observations, Scannapieco et al. (2006) have shown that, at $z\sim3$, 
the longitudinal C~{\sc iv} correlation function is consistent with a model 
where C~{\sc iv} is confined within bubbles of typical radius $\sim$2~Mpc comoving 
surrounding halos of mass $\sim$~10$^{12}$~M$_{\odot}$.
At this redshift, this corresponds to a separation of about $\sim$~2~arcmin. 
Unfortunately, the small size of our sample prevents any attempt to
consider only small separations. Our result is, however,  consistent
with their findings. 
\par\noindent
\subsection{Peculiarities}
It is interesting to note that there is an overdensity of C~{\sc iv} pairs in front of 
the quartet Q~0103$-$294A\&B, Q~0102$-$2931 and Q~0102$-$293.
The density of C~{\sc iv} systems along the four lines of sight (6.4 per unit redshift)
and the number of coincidences within 4000~km~s$^{-1}$ is about twice
larger than the mean density  of coincidences in the overall sample. 
In Fig.~\ref{f:Spectra}  we plot two portions of the spectra of
Q0103$-$294A,B (separated by 1.3~arcmin) and
Q~0102$-$2931 and Q~0102$-$293 (separated by about 6.5~arcmin from
Q0103$-$294A,B , see Table~1). The wavelength ranges of the two 
portions are centered on $z$~=~1.92 and 2.14, respectively. Note, however,
that the overdensity of C~{\sc iv} systems along these
lines of sight extends over the much larger redshift range
1.5~$\leq$~$z$~$\leq$~2.2 and over a spatial scale larger than 10~arcmin (see Appendix B). 
There are two more  peculiarities occurring  along these lines of sight.
There are  no C~{\sc iv} systems between $z$~=~1.955 and 2.051
and  there is a quasi-spherical structure of reduced H~{\sc i} 
absorption with radius $\sim$12.5$h^{-1}$~Mpc at $z$~$\sim$~1.992 
in front of the quartet (Rollinde et al. 2003). Note also that 
the correlation between the Lyman-$\alpha$ forests of Q~0102$-$2931 
and Q~0103$-$294B, two quasars of the group with $\sim$6~arcmin separation, 
is measured to be quite high ($\xi$~=~0.11, see Table~1 and Fig.~3).
\par\noindent
A similar overdensity of C~{\sc iv} systems has  been observed in the field 
of Tol~1037$-$2704 (e.g. Jakobsen et al. 1986; Dinshaw \& Impey 1996;
Lespine \& Petitjean 1997). The  overdensity of C~{\sc iv} systems 
in this field extends over the redshift range $\sim$1.5$-$2.2 
and a transverse scale $>$~15~arcmin and  has been interpreted as
being due to the presence of a supercluster. The dimensions of this
supercluster would be at least 80 and 30$h^{-1}$~Mpc along and perpendicular
to the line of sight, respectively. To our knowledge no deep imaging
of this field exists. Another overdensity 
of C~{\sc iv} systems has been reported in front of
PKS~0237$-$233 (Sargent, Boksenberg \& Steidel 1988; Foltz et al. 1993).
The overdensity reported in this work in the field around Q~0103$-$294A,B 
may give new clues to solve the puzzle of the origin of these 
overdensities extending over very large scales as the four quasars
constituting the quartet are very close to each other. Deep infra-red imaging
should be performed in the field to search for any concentration of objects
in the corresponding redshift range. High spectral resolution observations
of the quasars would allow a more detailed investigation of the nature
of these C~{\sc iv} systems.
\par\noindent
The pair J~000852.7-290044/J000857.7-290126 is also peculiar as the two lines of
sight show 9 and 4 C~{\sc iv} systems respectively, corresponding to
3.5 and 1.5 times the mean density of systems. J~000852.7-290044 shows a BAL 
systems and it would be interesting to question the intervening origin of some
of the narrow systems (Srianand \& Petitjean 2001).

\section{Conclusions}
We have obtained VLT-FORS observations of a large sample of 32 pairs of QSO
with separations in the range 0.6~$<$~$\theta$~$<$~10~arcmin  building 
on the smaller sample of Rollinde et al. (2003) of 11 pairs. We present 
measurements of the transverse and longitudinal correlation functions from 
this sample. We further use a large box-size DM only simulation and a somewhat 
smaller full hydro-dynamical simulation  to investigate the effect of spectral
resolution, thermal broadening and peculiar motions on the  correlation
function and to determine realistic error estimates. 
The  longitudinal correlation function from the FORS sample
is in good agreement with that obtained from UVES high-resolution data if
the effect of  the different spectral resolutions is taken into account. 

The transverse correlation is detected at the 
3~$\sigma$ level up to separations of about  $\sim$3$-$5~arcmin.
The sample is sufficiently large to obtain a binned
estimate of the average correlation function which has about a factor
1.5-2 smaller errors than the smaller sub-sample described in Rollinde
et al. (2003). The shape and correlation length of the transverse
correlation function of the absorbing gas is in good agreement
with expectations for absorption by density fluctuations in the warm 
photo-ionized Intergalactic Medium as described in CDM-like structure formation models.  
Our measurement of the transverse correlation function is
thus an important further independent confirmation that the  Lyman-$\alpha$
forest is  indeed caused by the filamentary and sheet-like structures
of the cosmic web predicted by these models.

We then use the numerical simulations and predictions of linear theory to
assess prospects of using the transverse correlation function for 
a variant of the  Alcock \& Paczy\'nski test  to determine cosmological 
parameters. In agreement with predictions of previous theoretical
studies we find that our sample is still too small for this purpose. 
The improved errors of our larger sample compared to the sub-sample 
of Rollinde et al. (2003) suggest however that meaningful constraints 
on $\Omega_{\Lambda}$ can be obtained. For this, a larger sample and a 
careful analysis of the systematic uncertainties 
with a large suite of full hydrodynamical simulations are necessary.
Mc~Donald (2003) estimated that this requires   
a sample of 13($\theta$/1')$^2$ pairs  on scales up to 10~arcmin. 

We have also used our sample to investigate the transverse and
longitudinal correlation functions of C~{\sc iv} absorption systems  
on the scales probed by our pairs, but did not detect any signal.  
This is not surprizing as most of the separations are larger than 2 
comoving Mpc. This is larger than is expected for the size of  
metal-enriched bubbles surrounding massive haloes
(e.g. Scannapieco et al. 2006).  
We have, however, detected a prominent overdensity of C~{\sc iv}
systems in front of the quartet Q~0103$-$294A\&B, Q~0102$-$2931 
and Q~0102$-$293 which extends  over the redshift range
1.5~$\leq$~$z$~$\leq$~2.2 and over a spatial scale larger than
10~arcmin. This suggests the presence of a high-redshift cluster 
in this field and makes it a prime target for deep infra-red imaging. 

\begin{table}
 \begin{center}
\begin{tabular}{|c c c c|}
 \hline
QSO & $w_{\rm obs}$ &$\lambda_{\rm{obs}}$& $z_{\rm abs}$ \\
    &  (\AA)        & (\AA)              &               \\
 \hline
J~135003.0$-$011703     & 29.6  & 4055.77 & 2.336  \\ 
J~144245.6$-$024251     & 21.4  & 3911.52 & 2.218  \\ 
J~000852.7$-$290044     & 22.2  & 3955.77 & 2.254  \\ 
J~000857.7$-$290126     & 20.0:  & 4243.08 & 2.490  \\
Q~2139$-$4504B          & 55.3  & 5071.54 & 3.172  \\
\hline
   \end{tabular}
 \end{center}
\caption{Damped Lyman-$\alpha$ system candidates
  detected in the survey }
\label{t:DLA}
\end{table}

%

%
\section*{Acknowledgements}
We thank D. Weinberg for useful discussions and for providing the numerical tables
of heating and photoionization rates used in the hydrodynamic simulation presented
in this paper. We thank an anonymous referee for a thorough reading of the manuscript
and detailed comments which significantly improved the paper.
The simulations were performed as part of a Numerical Investigations
in Cosmology group task  in the framework of the HORIZON project. 
Computer time for the simulations
was allocated by the scientific council of IDRIS, Orsay. FC thanks IUCAA-Pune (India)
for hospitality during the time part of this work has been completed and ESO-Vitacura
for a PhD studentship.

\vskip 8.cm

\appendix

\section{Comments on individual lines of sight}
In  this Section,  we comment on peculiarities of individual observed lines of sight.
The quasar emission redshifts (given in \Tab{pairs}) are determined by fitting a 
Gaussian profile to the C~{\sc iv} emission line when present in the spectrum or to the
Lyman-$\alpha$ emission line otherwise. Damped Lyman-$\alpha$  systems are listed in 
Table~2 and identified metal lines are given in Tables gathered in the Appendix.
\subsection{J~000852.7-290044$-$J~000857.7-290126}
QSO J~000852.7-290044 has a BAL system close to the emission redshift
with two strong components seen in N~{\sc v}, O~{\sc vi} and
C~{\sc iv}. In addition there is a damped Lyman-$\alpha$ system at \zabs~=~2.254 
($W_{\rm obs}$ = 22.2~\AA) with strong associated metallic absorption. There is no corresponding 
absorption toward J~000857.7-290126 but another damped Lyman-$\alpha$ system is
detected at \zabs~=~2.490 
($W_{\rm obs}$ = 20.0 \AA) with O~{\sc vi} and O~{\sc i} associated absorption. 
\cIV~ absorption is detected at \zabs~=~2.218 toward J~000852.7-290044 and at \zabs~=~2.215 
toward J~000857.7-290126 (i.e. with a velocity difference of only 294~\kms).
\subsection{Q~0103-294A$-$Q~0103-294B}
These quasars belong to a group of QSOs described in Rollinde et al. (2003). 
There is an over-density of \cIV systems between \zabs~=~1.536 and 2.18 observed
in front of the group
(see also Section~6.2). 
Alternative names for Q~0103-294A and B are,
respectively,  J010534.7-290917 and J010538.3-291106.
\subsection{Q~0236-2411$-$Q~0236-2413}
There are strong but narrow associated absorption features toward Q~0236-2413 close to the 
\cIV, Si~{\sc iv}  and N~{\sc v} emission lines.
\subsection{J~023836.9-282310$-$J~023849.0-282101}
We detect strong metallic absorptions toward QSO J~023836.9-282310 and in particular 
a strong O~{\sc vi} doublet at $z_{\rm abs}$~=~2.56.  
The signal to noise ratio of the J~023849.0-282101 
spectrum is good but very few metal absorptions are detected except for a strong
Mg~{\sc ii} system at \zabs~=~0.871. 
In the same spectrum a \cIV system may be present at \zabs~=~2.083 but the $\lambda$1550 transition 
is under our 3$\sigma$ detection limit.
\subsection{UM~680$-$UM~681}
This pair has been observed with UVES (see D'Odorico et al. 2002).
As discussed by these authors, there is a sub-DLA system (log~$N$(H~{\sc i})~=~18.6) 
toward UM~681 at $z_{\rm abs}$~=~1.788. Noticeably as well, there are two coincident 
Lyman Limit Systems at $z_{\rm abs}$~=~2.03 and two coincident associated systems at $z_{\rm abs}$~=~2.125.
\subsection{J~031036.4-305108$-$J~031041.0-305027}
Very few metal absorptions are seen toward J~031036.4-305108 apart from a \cIV system 
at \zabs~=~1.8. On the contrary, the line of sight toward J~031041.0-305027 shows two 
metallic systems, one at \zabs~$\sim$~2.39 and the other clearly associated with the 
QSO at \zabs~$\sim$~2.542. The two lines of the Si~{\sc iv} doublet for the latter system are under 
our 3$\sigma$ detection limit but weak features can be seen at the expected positions and
N~{\sc v} absorption lines are clearly detected (see Appendix). 
\subsection{FOCAP QSF:01$-$FOCAP QSF:04}
There is a possible \cIV doublet at \zabs~=~2.27 toward FOCAP QSF:01 but the 
corresponding C~{\sc iv}$\lambda$1550 transition is weaker than our detection limit.
Other names for FOCAP QSF:01 and FOCAP QSF:04 are, respectively,
J034105.1-445619 and J034126.2-445842.
\subsection{J~095800.2-002858$-$J~095810.9-002733}
Strong C~{\sc iv} absorption is seen toward J~095810.9-002733 at \zabs~$\sim$~1.807 with other 
metal absorption lines from Al~{\sc ii}, Al~{\sc iii} and C~{\sc ii};
the \lya\ absorption corresponding to this system is out of the observed wavelength
range. Possible N~{\sc v}$\lambda\lambda$1238,1242 absorptions are seen at 4164.5 and 
4151.5~\AA~ but the corresponding features are below the 3$\sigma$ detection limit. A Si~{\sc iv} 
system may be present at \zabs~=~2.122 toward J~095800.2-002858.   
\subsection{J~102827.1-013641$-$J~102832.6-013448}
The determination of the emission redshift for J~102827.1-013641 is complicated by the presence
of a number of absorptions at a redshift close to the emission redshift: there is 
a strong \lya\ system with associated metallic absorption at \zabs~=~2.399.
The determination of $z$ by a Gaussian fit of the 
\cIV emission line gives \zem~=~2.392. The Fe~{\sc ii} system at \zabs~=~1.316 has no
Mg~{\sc ii} counterpart detected.
\subsection{Q~1206-1056$-$Q~1207-1057}
Q~1207-1057 shows broad and shallow absorptions at 4885-4955~\AA~  for \cIV, 4385-4500~\AA~  for 
S~{\sc iv} and 3900-3974~\AA~  for N~{\sc v}. The \lya\ line associated with the BAL 
system is not clearly detected. 
There is probably a Mg~{\sc ii} and Fe~{\sc ii} system at $z_{\rm abs}$~$\sim$~0.772
but most of the corresponding absorptions are under the 3$\sigma$ detection limit. 
\subsection{J~120725.9-024519$-$J~120734.5-024725}
Strong absorptions from Mg~{\sc ii}$\lambda\lambda$2796,2803, 
Fe~{\sc ii}$\lambda\lambda$2374,2382,
Fe~{\sc ii}$\lambda\lambda$2596,2600
and Mg{\sc i}$\lambda$2852 are detected at \zabs~=~0.777 toward J~120725.9-024519. 
\subsection{J~123510.5-010746$-$J~123511.0-010830}
The redshift difference between these two quasars is one of the largest in our sample:
$z_{\rm em}$~=~2.785 and 2.235 for J~123510.5-010746 and J~123511.0-010830
respectively.
There is a $z_{\rm abs}\sim2.26$ associated \cIV system in the spectrum of J~123510.5-010746 at a 
redshift close to the emission redshift of the QSO. 
\subsection{J~125556.9+001848$-$J~125606.3+001728}
The spectrum of J~125556.9+001848 presents a shallow $z_{\rm abs}\sim2.08$ \cIV absorption feature close to 
the emission redshift and a strong absorption in the range 3810-3845~\AA~ that could be identified as 
the corresponding N~{\sc v} absorption.
J~125606.3+001728 shows several strong \cIV absorptions 
that have no counterpart along the adjacent line of sight toward J~125556.9+001848. 
%
%
\subsection{J~135001.7-011703$-$J~135003.0-011819}
A damped Lyman-$\alpha$ system is detected toward J~135003.0-011819 at \zabs~=~2.33 
($W_{\rm obs}$ = 29.6 \AA) with associated strong metallic absorption.
\subsection{J~141124.6-022943$-$J~141117.3-023222}
There is a noticeable decrease of the number of H~{\sc i} absorption lines in the \lya\ forest 
of J~141124.6-022943 at the emission redshift of J~141117.3-023222 possibly corresponding to
a strong transverse proximity effect.
%
%
%
\subsection{J144245.6-024251$-$J144245.7-023906}
Toward J~144245.6-024251, there is a damped Lyman-$\alpha$ system at \zabs~=~2.218 
($W_{\rm obs}$~=~21.4 \AA) as well as Zn~{\sc ii}, Cr~{\sc ii} and  Fe~{\sc ii} absorptions at 
\zabs~=~1.178.
\subsection{Q~2129-4653A$-$Q~2129-4653B}
There is a strong feature in the two spectra over the wavelength range 4044-4058~\AA. 
In Q2129-4653B we successfully identified this feature as two blended \cIV systems
with a separation of about 530 \kms. Along the other line of sight the lines 
are heavily blended but could be modelled as C~{\sc iv} absorptions 
at the same redshift.
\subsection{Q~2139-4504B$-$Q~2139-4504A}
This is the pair with the smallest separation (0.6 arcmin) and the highest redshift 
in our sample. The two spectra show a Lyman limit system at a redshift close to
the emission redshift of the quasar. 
There is a strong damped Lyman-$\alpha$ system at \zabs~=~3.172 ($W_{\rm obs}$ = 55.35 \AA) 
toward Q~2139-4504B with associated C~{\sc ii} and Si~{\sc ii} absorptions. 
There are no corresponding metal absorptions toward Q~2139-4504A down to
$w_{\rm r}$~$<$~0.3~\AA. It is interesting to note that there is a lack of absorption 
in the \lya~ forest of Q~2139-4504B at the redshift of Q~2139-4504A suggesting 
the presence of a strong transverse proximity effect.
\subsection{J~214501.6-303121$-$J~214507.0-303046}
Associated systems are detected toward both quasars, at $-$1250 and $-$380~km~s$^{-1}$ 
relative to the QSO emission redshift toward,
respectively, J~214501.6-303121 and J~214507.0-303046.
\subsection{J~223850.1-295612$-$J~223850.9-295301}
J~223850.9-295301 exhibits broad but shallow absorption lines of \cIV and \lya.   
%
%
%
\subsection{J~232800.7-271655$-$J~232804.4-271713}
The spectra have a poor signal-to-noise ratio.
We identify a possible Mg~{\sc ii} system toward J~232800.7-271655 at \zabs~=~0.368 
(Mg~{\sc ii}$\lambda$2803 
and Mg~{\sc ii}$\lambda$2796)
but no other species are seen in the spectrum. 
A strong \cIV system may be present in the \lya\ forest of J~232804.4-271713 
at \zabs~=~1.545.
%
%
%
%
%
%
%
\subsection{Q0102-293 ~~\zem~=~2.441  and Q0102-2931 ~~\zem~=~2.212}
Alternative names for Q~0102-293 and Q~0102-2931 are,
respectively, J010502.8-290618 and J010518.0-291510.
\par\noindent
\par\noindent
\vfill
\newpage

\section{Line Lists}
\newpage
\begin{table}
\caption{Line list for J~000852.7-290044 and J~000857.7-290126}
 \begin{tiny}
 \begin{center}
\begin{tabular}{|p{0.0005cm} p{0.2cm} p{0.1cm} p{0.4cm} p{0.7cm} p{0.33cm} |  p{0.2cm} p{0.1cm} p{0.4cm} p{0.7cm} p{0.33cm}|}
 \hline
 & \multicolumn{3}{c}{J~000852.7-290044} & & & & \multicolumn{3}{c}{J~000857.7-290126}&\\
 &$w_{\rm{obs}}$&S/N&$\lambda_{\rm{obs}}$&Ident&z&$w_{\rm{obs}}$&S/N&$\lambda_{\rm{obs}}$&Ident&z \\
 \hline
1&11.21& 16& 3752.73& OVI\_1031  & 2.6366&$<$0.37 & && & \\ 
2&7.25& 18& 3758.33& OVI\_1031  & 2.6420&$<$0.39 & && & \\ 
3&6.06& 21& 3773.53& OVI\_1037  & 2.6367&$<$0.51 & && & \\ 
4&9.25& 32& 3777.46& OVI\_1037  & 2.6405&$<$0.62 & && & \\ 
5&1.38& 56& 3872.01& SiII\_1190 & 2.2527&$<$1.74 & && & \\ 
6&1.36& 57& 3882.08& SiII\_1193 & 2.2533&$<$2.45 & && & \\
\hline 
\end{tabular}
 \end{center}
 \end{tiny}
\end{table}
\begin{table}
\caption{Line list for J~005852.4-272933 and J~005859.1-273038}
 \begin{tiny}
 \begin{center}
 \begin{tabular}{|p{0.0005cm} p{0.2cm} p{0.1cm} p{0.4cm} p{0.7cm} p{0.33cm} |  p{0.2cm} p{0.1cm} p{0.4cm} p{0.7cm} p{0.33cm}|}
 \hline
 & \multicolumn{3}{c}{J~005852.4-272933} & & & & \multicolumn{3}{c}{J~005859.1-273038}&\\
 &$w_{\rm{obs}}$&S/N&$\lambda_{\rm{obs}}$&Ident&z&$w_{\rm{obs}}$&S/N&$\lambda_{\rm{obs}}$&Ident&z \\
 \hline
1&$<$3.25 & & & & &2.44& 25& 4161.45& SiII\_1190 & 2.4958\\ 
2&$<$2.12 & & & & &2.63& 23& 4171.79& SiII\_1193 & 2.4960\\ 
3&$<$0.40 & & & & &2.73& 25& 4217.50& SiIII\_1206 & 2.4956\\ 
4&2.14& 29& 4277.63& CII\_1334  & 2.2053&$<$0.97 & && & \\ 
& & & & HI\_1215   & 2.519 & & && & \\ 
5&$<$0.30 & & & & &1.68& 43& 4406.39& SiII\_1260 & 2.4960\\ 
6&0.43& 36& 4457.53& ...& ...& & & & & \\ 
7&0.41& 34& 4468.29& SiIV\_1393 & 2.2059&$<$0.32 & && & \\ 
8&0.22& 34& 4496.46& SiIV\_1402 & 2.2054&$<$0.34 & && & \\ 
9&0.62& 33& 4508.74& ...& ...& & & & & \\ 
10&$<$0.41 & & & & &0.71& 30& 4560.23& SiII\_1304 & 2.4961\\ 
11& & & & & &0.41& 33& 4642.71& ...& ...\\ 
13&$<$0.38 & & & & &1.86& 31& 4665.65& CII\_1334  & 2.4961\\ 
14&0.67& 31& 4773.01& CIV\_1548  & 2.0829&$<$0.37 & && & \\ 
15&0.43& 32& 4778.46& CIV\_1550  & 2.0813&$<$0.38 & && & \\ 
16&$<$0.40 & & & & &2.33& 27& 4872.33& SiIV\_1393 & 2.4958\\ 
17&$<$0.40 & & & & &1.61& 29& 4903.67& SiIV\_1402 & 2.4957\\ 
18&0.67& 32& 4963.17& CIV\_1548  & 2.2058&$<$0.33 & && & \\ 
19&0.34& 34& 4972.17& CIV\_1550  & 2.2062&$<$0.32 & && & \\ 
20& & & & & &0.49& 39& 5006.94& ...& ...\\ 
21&0.36& 32& 5091.25& ...& ...& & & & & \\ 
22&$<$0.42 & & & & &0.90& 29& 5337.35& SiII\_1526 & 2.4960\\ 
23&0.25& 29& 5338.66& ...& ...& & & & & \\ 
24&$<$0.39 & & & & &4.02& 25& 5411.35& CIV\_1548  & 2.4952\\ 
25&$<$0.39 & & & & &2.97& 28& 5420.39& CIV\_1550  & 2.4953\\ 
26&0.78& 29& 5652.77& ...& ...& & & & & \\ 
27&0.96& 27& 5731.73& ...& ...& & & & & \\ 
 \hline 
 \end{tabular}
 \end{center}
 \end{tiny}
\end{table}
\begin{table}
\caption{Line list for Q0102-2931 and Q0102-293}
 \begin{tiny}
 \begin{center}
 \begin{tabular}{|p{0.0005cm} p{0.2cm} p{0.1cm} p{0.4cm} p{0.7cm} p{0.33cm} |  p{0.2cm} p{0.1cm} p{0.4cm} p{0.7cm} p{0.33cm}|}
 \hline
 & \multicolumn{3}{c}{Q0102-2931} & & & & \multicolumn{3}{c}{Q0102-293}&\\
 &$w_{\rm{obs}}$&S/N&$\lambda_{\rm{obs}}$&Ident&z&$w_{\rm{obs}}$&S/N&$\lambda_{\rm{obs}}$&Ident&z \\
 \hline
1&0.22& 41& 3943.28& CII\_1334: & 1.9548&$<$0.45 & && & \\ 
2&0.43& 27& 4114.44& SiIV\_1393 & 1.9520&$<$2.49 & && & \\ 
3&0.32& 28& 4117.41& SiIV\_1393 & 1.9542&$<$0.91 & && & \\ 
4&0.37& 27& 4140.48& SiIV\_1402 & 1.9516&$<$0.27 & && & \\ 
5&0.30& 28& 4144.54& SiIV\_1402 & 1.9545&$<$0.64 & && & \\ 
6&$<$0.43 & & & & &0.74& 48& 4228.74& CIV\_1548  & 1.7314\\
\hline 
\end{tabular}
 \end{center}
 \end{tiny}
\end{table}
\begin{table}
\caption{Line list for Q0103-294A and Q0103-294B}
 \begin{tiny}
 \begin{center}
 \begin{tabular}{|p{0.0005cm} p{0.2cm} p{0.1cm} p{0.4cm} p{0.7cm} p{0.33cm} |  p{0.2cm} p{0.1cm} p{0.4cm} p{0.7cm} p{0.33cm}|}
 \hline
 & \multicolumn{3}{c}{Q0103-294A} & & & & \multicolumn{3}{c}{Q0103-294B}&\\
 &$w_{\rm{obs}}$&S/N&$\lambda_{\rm{obs}}$&Ident&z&$w_{\rm{obs}}$&S/N&$\lambda_{\rm{obs}}$&Ident&z \\
 \hline
1&0.13& 63& 3904.00& ...& ...& & & & & \\ 
3&$<$0.20 & & & & &0.37& 49& 3925.59& CIV\_1548  & 1.5356\\ 
4&$<$0.20 & & & & &0.26& 48& 3932.93& CIV\_1550  & 1.5361\\ 
5&$<$0.20 & & & & &0.44& 47& 3941.42& CIV\_1548  & 1.5458\\ 
6&$<$0.20 & & & & &0.22& 49& 3947.88& CIV\_1550  & 1.5457\\ 
7&$<$0.20 & & & & &1.76& 43& 3956.64& CIV\_1548: & 1.5556\\
\hline 
\end{tabular}
 \end{center}
 \end{tiny}
\end{table}
\begin{table}
\caption{Line list for J~013734.2-303802 and J~013734.2-304050}
 \begin{tiny}
 \begin{center}
 \begin{tabular}{|p{0.0005cm} p{0.2cm} p{0.1cm} p{0.4cm} p{0.7cm} p{0.33cm} |  p{0.2cm} p{0.1cm} p{0.4cm} p{0.7cm} p{0.33cm}|}
  \hline
 & \multicolumn{3}{c}{J~013734.2-303802} & & & & \multicolumn{3}{c}{J~013734.2-304050}&\\
 &$w_{\rm{obs}}$&S/N&$\lambda_{\rm{obs}}$&Ident&z&$w_{\rm{obs}}$&S/N&$\lambda_{\rm{obs}}$&Ident&z \\
 \hline
1&$<$1.00 & & & & &0.33& 58& 4128.70& CII\_1334  & 2.0937\\ 
2& & & & & &0.33& 44& 4216.09& ...& ...\\ 
5&0.46& 40& 4291.65& ...& ...& & & & & \\ 
6&0.47& 41& 4321.63& NV\_1238   & 2.4885&$<$0.31 & && & \\ 
7&0.21& 40& 4335.65& NV\_1242   & 2.4886&$<$0.31 & && & \\ 
9&0.32& 37& 4353.71& ...& ...& & & & & \\ 
10&$<$0.40 & & & & &0.27& 39& 4500.24& SiIV\_1393 & 2.2289\\ 
11&$<$0.40 & & & & &0.35& 40& 4529.05& SiIV\_1402 & 2.2286\\ 
12&$<$0.41 & & & & &0.40& 44& 4636.20& CIV\_1548  & 1.9946\\ 
13&$<$0.40 & & & & &0.30& 45& 4643.51& CIV\_1550  & 1.9943\\ 
14&$<$0.40 & & & & &0.62& 39& 4790.69& CIV\_1548  & 2.0944\\ 
15&$<$0.39 & & & & &0.41& 40& 4798.57& CIV\_1550  & 2.0943\\ 
18&$<$0.36 & & & & &0.65& 39& 4849.10& CIV\_1548  & 2.1321\\ 
19&$<$0.36 & & & & &0.41& 39& 4856.68& CIV\_1550  & 2.1318\\ 
20&0.96& 33& 4861.62& SiIV\_1393 & 2.4881&$<$0.31 & && & \\ 
21&0.22& 34& 4873.05& ...& ...& & & & & \\ 
22&0.54& 33& 4892.82& SiIV\_1402 & 2.4880&$<$0.31 & && & \\ 
23&$<$0.40 & & & & &1.16& 40& 4999.07& CIV\_1548  & 2.2289\\ 
24&$<$0.40 & & & & &0.89& 41& 5007.56& CIV\_1550  & 2.2291\\ 
25&0.32& 32& 5239.63& ...& ...& & & & & \\ 
26&2.34& 50& 5400.30& CIV\_1548  & 2.4881&$<$0.29 & && & \\ 
27&1.86& 48& 5409.27& CIV\_1550  & 2.4881&$<$0.29 & && & \\ 
 \hline 
 \end{tabular}
 \end{center}
 \end{tiny}
\end{table}
\begin{table}
\caption{Line list for Q0236-2411 and Q0236-2413}
 \begin{tiny}
 \begin{center}
 \begin{tabular}{|p{0.0005cm} p{0.2cm} p{0.1cm} p{0.4cm} p{0.7cm} p{0.33cm} |  p{0.2cm} p{0.1cm} p{0.4cm} p{0.7cm} p{0.33cm}|}
 \hline
 & \multicolumn{3}{c}{Q0236-2411} & & & & \multicolumn{3}{c}{Q0236-2413}&\\
 &$w_{\rm{obs}}$&S/N&$\lambda_{\rm{obs}}$&Ident&z&$w_{\rm{obs}}$&S/N&$\lambda_{\rm{obs}}$&Ident&z \\
 \hline
1&$<$1.37 & & & & &5.87& 25& 3947.47& NV\_1238   & 2.1865\\ 
2&$<$0.60 & & & & &6.31& 24& 3960.15& NV\_1242   & 2.1865\\ 
3&$<$1.30 & & & & &2.69& 34& 3971.02& NV\_1238   & 2.2055\\ 
4&$<$1.44 & & & & &1.72& 38& 3984.24& NV\_1242   & 2.2058\\ 
6& & & & & &0.53& 35& 4007.95& ...& ...\\ 
7&0.82& 41& 4046.73& NV\_1238   & 2.2666&$<$0.38 & && & \\
\hline 
\end{tabular}
 \end{center}
 \end{tiny}
\end{table}
\begin{table}
\caption{Line list for J~023836.9-282310 and J~023849.0-282101}
 \begin{tiny}
 \begin{center}
 \begin{tabular}{|p{0.0005cm} p{0.2cm} p{0.1cm} p{0.4cm} p{0.7cm} p{0.33cm} |  p{0.2cm} p{0.1cm} p{0.4cm} p{0.7cm} p{0.33cm}|}
 \hline
 & \multicolumn{3}{c}{J~023836.9-282310} & & & & \multicolumn{3}{c}{J~023849.0-282101}&\\
 &$w_{\rm{obs}}$&S/N&$\lambda_{\rm{obs}}$&Ident&z&$w_{\rm{obs}}$&S/N&$\lambda_{\rm{obs}}$&Ident&z \\
 \hline
1&4.02& 22& 3671.50& OVI\_1031: & 2.5579&$<$0.28 & && & \\ 
2&3.50& 28& 3691.45& OVI\_1037: & 2.5576&$<$0.24 & && & \\ 
4& & & & & &0.28& 84& 4385.94& ...& ...\\ 
5&0.53& 59& 4386.18& FeII\_2382 & 0.8408&$<$0.25 & && & \\ 
6&4.61& 50& 4406.52& SiIV\_1393 & 2.1616&$<$0.15 & && & \\ 
7&4.61& 50& 4406.52& NV\_1238   & 2.5570&$<$0.15 & && & \\
\hline 
 \end{tabular}
 \end{center}
 \end{tiny}
\end{table}

\begin{table}
\caption{Line list for J~031036.4-305108 and J~031041.0-305027}
 \begin{tiny}
 \begin{center}
 \begin{tabular}{|p{0.0005cm} p{0.2cm} p{0.1cm} p{0.4cm} p{0.7cm} p{0.33cm} |  p{0.2cm} p{0.1cm} p{0.4cm} p{0.7cm} p{0.33cm}|}
 \hline
 & \multicolumn{3}{c}{J~031036.4-305108} & & & & \multicolumn{3}{c}{J~031041.0-305027}&\\
 &$w_{\rm{obs}}$&S/N&$\lambda_{\rm{obs}}$&Ident&z&$w_{\rm{obs}}$&S/N&$\lambda_{\rm{obs}}$&Ident&z \\
 \hline
1&1.11& 17& 3736.42& CII\_1334  & 1.7998&$<$0.62 & && & \\ 
& & & & HI\_1215   & 2.074 & & && & \\ 
2&0.68& 52& 4334.83& CIV\_1548  & 1.7999&$<$0.20 & && & \\ 
3&0.78& 48& 4341.97& CIV\_1550  & 1.7999&$<$0.21 & && & \\ 
4&0.28& 40& 4380.07& ...& ...& & & & & \\ 
5&$<$0.31 & & & & &0.81& 51& 4388.24& NV\_1238   & 2.5423\\ 
6&0.48& 38& 4393.47& NV\_1238:  & 2.5465&$<$0.24 & && & \\ 
7&$<$0.32 & & & & &0.47& 49& 4402.48& NV\_1242   & 2.5424\\ 
8&0.54& 38& 4407.88& NV\_1242:  & 2.5467&$<$0.25 & && & \\ 
9& & & & & &0.28& 48& 4412.16& ...& ...\\ 
12&$<$0.43 & & & & &0.65& 33& 4649.20& MgII\_2796 & 0.6626\\ 
13&$<$0.45 & & & & &0.62& 33& 4661.19& MgII\_2803 & 0.6626\\ 
14&0.37& 26& 4677.29& AlII\_1670 & 1.7995&$<$0.37 & && & \\ 
15&$<$0.47 & & & & &1.12& 32& 4725.06& SiIV\_1393 & 2.3902\\ 
16&$<$0.46 & & & & &0.57& 33& 4754.80& SiIV\_1402 & 2.3896\\ 
17&0.77& 26& 5079.98& ...& ...& & & & & \\ 
18&0.85& 26& 5193.08& AlIII\_1854 & 1.7999&$<$0.39 & && & \\ 
19&0.69& 27& 5215.09& AlIII\_1862 & 1.7996&$<$0.38 & && & \\ 
20&$<$0.45 & & & & &2.89& 27& 5247.96& CIV\_1548  & 2.3897\\ 
21&$<$0.45 & & & & &2.35& 26& 5256.67& CIV\_1550  & 2.3897\\ 
22&$<$0.31 & & & & &0.94& 46& 5484.23& CIV\_1548  & 2.5423\\ 
23&$<$0.70 & & & & &0.95& 46& 5494.10& CIV\_1550  & 2.5428\\ 
24&1.13& 42& 5499.14& CIV\_1548  & 2.5519&$<$0.51 & && & \\ 
25&0.32& 45& 5507.67& CIV\_1550  & 2.5515&$<$0.26 & && & \\ 
 \hline 
\end{tabular}
 \end{center}
 \end{tiny}
\end{table}
\begin{table}
\caption{Line list for J~031054.7-293436 and J~031103.0-293306}
 \begin{tiny}
 \begin{center}
 \begin{tabular}{|p{0.0005cm} p{0.2cm} p{0.1cm} p{0.4cm} p{0.7cm} p{0.33cm} |  p{0.2cm} p{0.1cm} p{0.4cm} p{0.7cm} p{0.33cm}|}
 \hline
 & \multicolumn{3}{c}{J~031054.7-293436} & & & & \multicolumn{3}{c}{J~031103.0-293306}&\\
 &$w_{\rm{obs}}$&S/N&$\lambda_{\rm{obs}}$&Ident&z&$w_{\rm{obs}}$&S/N&$\lambda_{\rm{obs}}$&Ident&z \\
 \hline
1&$<$0.50 & & & & &0.98& 59& 3892.20& CIV\_1548  & 1.5140\\ 
2&$<$0.73 & & & & &0.68& 58& 3898.58& CIV\_1550  & 1.5139\\ 
4&0.52& 34& 4073.00& ...& ...& & & & & \\ 
5&0.68& 35& 4081.17& NV\_1238   & 2.2944&$<$0.29 & && & \\ 
6&0.36& 34& 4094.45& NV\_1242   & 2.2945&$<$0.29 & && & \\ 
7&0.53& 31& 4115.50& ...& ...& & & & & \\
\hline 
 \end{tabular}
 \end{center}
 \end{tiny}
\end{table}
\begin{table}
\caption{Line list for FOCAP QSF:01 and FOCAP QSF:04}
 \begin{tiny}
 \begin{center}
 \begin{tabular}{|p{0.0005cm} p{0.2cm} p{0.1cm} p{0.4cm} p{0.7cm} p{0.33cm} |  p{0.2cm} p{0.1cm} p{0.4cm} p{0.7cm} p{0.33cm}|}
 \hline
 & \multicolumn{3}{c}{FOCAP QSF:01} & & & & \multicolumn{3}{c}{FOCAP QSF:04}&\\
 &$w_{\rm{obs}}$&S/N&$\lambda_{\rm{obs}}$&Ident&z&$w_{\rm{obs}}$&S/N&$\lambda_{\rm{obs}}$&Ident&z \\
 \hline
1&1.65& 19& 3791.80& FeII\_2382 & 0.5913&$<$0.38 & && & \\ 
2&$<$1.37 & & & & &0.52& 29& 3823.11& SiIV\_1393 & 1.7430\\ 
3&$<$0.48 & & & & &0.48& 28& 3848.52& SiIV\_1402 & 1.7435\\ 
4&3.10& 34& 3930.54& AlII\_1670: & 1.3525&$<$0.46 & && & \\ 
7&0.39& 45& 4044.94& NV\_1238   & 2.2652&$<$0.43 & && & \\ 
8&$<$0.27 & & & & &1.43& 26& 4054.02& CIV\_1548  & 1.6185\\
\hline 
 \end{tabular}
 \end{center}
 \end{tiny}
\end{table}
\begin{table}
\caption{Line list for J~095800.2-002858 and J~095810.9-002733}
 \begin{tiny}
 \begin{center}
 \begin{tabular}{|p{0.0005cm} p{0.2cm} p{0.1cm} p{0.4cm} p{0.7cm} p{0.33cm} |  p{0.2cm} p{0.1cm} p{0.4cm} p{0.7cm} p{0.33cm}|}
 \hline
 & \multicolumn{3}{c}{J~095800.2-002858} & & & & \multicolumn{3}{c}{J~095810.9-002733}&\\
 &$w_{\rm{obs}}$&S/N&$\lambda_{\rm{obs}}$&Ident&z&$w_{\rm{obs}}$&S/N&$\lambda_{\rm{obs}}$&Ident&z \\
 \hline
1&$<$1.69 & & & & &5.47& 10& 3745.16& CII\_1334  & 1.8063\\ 
& & & & & & & &  & HI\_1215   & 2.0807\\ 
2&$<$0.69 & & & & &2.83& 16& 3937.55& SiIV\_1402 & 1.8070\\ 
& & & & & & & &  & HI\_1215   & 2.2390\\ 
3&2.22& 31& 4144.25& CIV\_1548  & 1.6768&$<$3.98 & && & \\ 
4&2.24& 30& 4151.67& CIV\_1550  & 1.6771&$<$1.03 & && & \\
\hline 
\end{tabular}
 \end{center}
 \end{tiny}
\end{table}
\begin{table}
\caption{Line list for J~102827.1-013641 and J~102832.6-013448}
 \begin{tiny}
 \begin{center}
 \begin{tabular}{|p{0.0005cm} p{0.2cm} p{0.1cm} p{0.4cm} p{0.7cm} p{0.33cm} |  p{0.2cm} p{0.1cm} p{0.4cm} p{0.7cm} p{0.33cm}|}
 \hline
 & \multicolumn{3}{c}{J~102827.1-013641} & & & & \multicolumn{3}{c}{J~102832.6-013448}&\\
 &$w_{\rm{obs}}$&S/N&$\lambda_{\rm{obs}}$&Ident&z&$w_{\rm{obs}}$&S/N&$\lambda_{\rm{obs}}$&Ident&z \\
 \hline
1&$<$0.90 & & & & &1.09& 21& 3878.17& CII\_1334  & 1.9060\\ 
& & & & & & & &  & HI\_1215   & 2.1901\\ 
2&2.21& 15& 4046.38& SiII\_1190 & 2.3991&$<$0.57 & && & \\ 
3&$<$0.79 & & & & &0.97& 35& 4050.95& SiIV\_1393 & 1.9065\\ 
4&0.60& 17& 4055.78& SiII\_1193 & 2.3988&$<$0.33 & && & \\ 
5&$<$1.19 & & & & &0.73& 37& 4076.75& SiIV\_1402 & 1.9062\\
\hline 
 \end{tabular}
 \end{center}
 \end{tiny}
\end{table}

\begin{table}
\caption{Line list for J~111200.4-013242 and J~111201.8-013018}
 \begin{tiny}
 \begin{center}
 \begin{tabular}{|p{0.0005cm} p{0.2cm} p{0.1cm} p{0.4cm} p{0.7cm} p{0.33cm} |  p{0.2cm} p{0.1cm} p{0.4cm} p{0.7cm} p{0.33cm}|}
 \hline
 & \multicolumn{3}{c}{J~111200.4-013242} & & & & \multicolumn{3}{c}{J~111201.8-013018}&\\
 &$w_{\rm{obs}}$&S/N&$\lambda_{\rm{obs}}$&Ident&z&$w_{\rm{obs}}$&S/N&$\lambda_{\rm{obs}}$&Ident&z \\
 \hline
1&$<$0.81 & & & & &1.64& 43& 3933.49& SiIV\_1393 & 1.8222\\ 
2&$<$1.72 & & & & &1.06& 50& 3959.22& SiIV\_1402 & 1.8224\\ 
3&0.24& 50& 4127.70& ...& ...& & & & & \\ 
4&0.32& 45& 4151.18& CIV\_1548  & 1.6813&$<$0.74 & && & \\ 
5&0.48& 46& 4157.79& CIV\_1550  & 1.6811&$<$0.21 & && & \\ 
6&0.27& 39& 4194.57& CIV\_1548  & 1.7093&$<$0.40 & && & \\
\hline 
\end{tabular}
 \end{center}
 \end{tiny}
\end{table}
\begin{table}
\caption{Line list for J~112108.2+003420 and J~112116.1+003112}
 \begin{tiny}
 \begin{center}
 \begin{tabular}{|p{0.0005cm} p{0.2cm} p{0.1cm} p{0.4cm} p{0.7cm} p{0.33cm} |  p{0.2cm} p{0.1cm} p{0.4cm} p{0.7cm} p{0.33cm}|}
 \hline
 & \multicolumn{3}{c}{J~112108.2+003420} & & & & \multicolumn{3}{c}{J~112116.1+003112}&\\
 &$w_{\rm{obs}}$&S/N&$\lambda_{\rm{obs}}$&Ident&z&$w_{\rm{obs}}$&S/N&$\lambda_{\rm{obs}}$&Ident&z \\
 \hline
1&0.32& 46& 3935.02& ...& ...& & & & & \\ 
2& & & & & &0.51& 31& 4019.10& ...& ...\\ 
3&0.40& 36& 4065.07& ...& ...& & & & & \\ 
4&0.58& 34& 4085.84& CII\_1334  & 2.0616&$<$0.42 & && & \\ 
5& & & & & &0.65& 27& 4125.55& ...& ...\\ 
6& & & & & &0.58& 25& 4246.95& ...& ...\\ 
7&1.23& 34& 4267.73& SiIV\_1393 & 2.0620&$<$1.32 & && & \\ 
8& & & & & &1.89& 26& 4279.14& ...& ...\\ 
9&1.14& 33& 4295.44& SiIV\_1402 & 2.0621&$<$0.44 & && & \\ 
10& & & & & &0.31& 30& 4464.20& ...& ...\\ 
11&0.53& 38& 4492.89& ...& ...& & & & & \\ 
12&$<$0.34 & & & & &1.00& 27& 4553.79& CIV\_1548  & 1.9413\\ 
13&$<$0.34 & & & & &0.68& 27& 4561.16& CIV\_1550  & 1.9412\\ 
14&0.66& 35& 4660.61& ...& ...& & & & & \\ 
15&$<$0.35 & & & & &0.85& 26& 4668.46& CIV\_1548  & 2.0154\\ 
16&$<$0.35 & & & & &0.70& 27& 4675.78& CIV\_1550  & 2.0151\\ 
17&$<$0.34 & & & & &1.46& 26& 4717.58& CIV\_1548  & 2.0471\\ 
18&$<$0.34 & & & & &1.16& 27& 4725.59& CIV\_1550  & 2.0472\\ 
19&1.54& 34& 4741.18& CIV\_1548  & 2.0624&$<$0.43 & && & \\ 
20&1.51& 33& 4748.41& CIV\_1550  & 2.0619&$<$0.43 & && & \\ 
21& & & & & &0.44& 28& 4793.89& ...& ...\\ 
22&$<$0.27 & & & & &1.97& 46& 4976.65& CIV\_1548  & 2.2145\\ 
23&$<$0.28 & & & & &0.97& 43& 4985.13& CIV\_1550  & 2.2146\\ 
25& & & & & &0.39& 30& 5374.31& ...& ...\\ 
26&$<$0.34 & & & & &0.86& 28& 5407.04& MgII\_2796: & 0.9336\\ 
27&$<$0.34 & & & & &0.50& 28& 5419.57& MgII\_2803: & 0.9331\\ 
 \hline 
\end{tabular}
 \end{center}
 \end{tiny}
\end{table}
\begin{table}
\caption{Line list for J~120725.9-024519 and J~120734.5-024725}
 \begin{tiny}
 \begin{center}
 \begin{tabular}{|p{0.0005cm} p{0.2cm} p{0.1cm} p{0.4cm} p{0.7cm} p{0.33cm} |  p{0.2cm} p{0.1cm} p{0.4cm} p{0.7cm} p{0.33cm}|}
 \hline
 & \multicolumn{3}{c}{J~120725.9-024519} & & & & \multicolumn{3}{c}{J~120734.5-024725}&\\
 &$w_{\rm{obs}}$&S/N&$\lambda_{\rm{obs}}$&Ident&z&$w_{\rm{obs}}$&S/N&$\lambda_{\rm{obs}}$&Ident&z \\
 \hline
1&$<$1.10 & & & & &0.26& 41& 3985.24& CII\_1334  & 1.9862\\ 
2& & & & & &0.76& 28& 4129.70& ...& ...\\ 
3& & & & & &0.91& 27& 4162.23& ...& ...\\ 
4&1.99& 31& 4235.65& FeII\_2382 & 0.7776&$<$0.42 & && & \\ 
5&$<$2.40 & & & & &1.74& 26& 4323.53& CIV\_1548  & 1.7926\\ 
6&$<$0.33 & & & & &1.67& 26& 4330.54& CIV\_1550  & 1.7925\\ 
7& & & & & &0.22& 32& 4528.50& ...& ...\\ 
8&1.56& 54& 4548.92& NV\_1238   & 2.6720&$<$0.36 & && & \\ 
9&1.25& 53& 4563.67& NV\_1242   & 2.6721&$<$0.36 & && & \\ 
10&0.59& 44& 4597.41& FeII\_2586 & 0.7774&$<$0.40 & && & \\ 
11&0.91& 38& 4622.01& FeII\_2600 & 0.7776&$<$1.25 & && & \\ 
12&$<$0.97 & & & & &1.86& 27& 4623.96& CIV\_1548  & 1.9867\\ 
13&$<$0.30 & & & & &1.19& 28& 4631.06& CIV\_1550  & 1.9863\\ 
14&0.58& 36& 4685.33& CIV\_1548: & 2.0263&$<$0.42 & && & \\ 
15&0.38& 36& 4693.78& CIV\_1550: & 2.0267&$<$0.42 & && & \\ 
17&1.53& 34& 4970.20& MgII\_2796 & 0.7774&$<$0.33 & && & \\ 
18&1.08& 33& 4983.03& MgII\_2803 & 0.7774&$<$0.32 & && & \\ 
21&1.65& 52& 5685.12& CIV\_1548  & 2.6721&$<$0.46 & && & \\ 
22&1.33& 53& 5694.60& CIV\_1550  & 2.6721&$<$0.45 & && & \\ 
 \hline 
 \end{tabular}
 \end{center}
 \end{tiny}
\end{table}

\begin{table}
\caption{Line list for Q~1206-1056 and Q~1207-1057}
 \begin{tiny}
 \begin{center}
 \begin{tabular}{|p{0.0005cm} p{0.2cm} p{0.1cm} p{0.4cm} p{0.7cm} p{0.33cm} |  p{0.2cm} p{0.1cm} p{0.4cm} p{0.7cm} p{0.33cm}|}
 \hline
 & \multicolumn{3}{c}{Q~1206-1056} & & & & \multicolumn{3}{c}{Q~1207-1057}&\\
 &$w_{\rm{obs}}$&S/N&$\lambda_{\rm{obs}}$&Ident&z&$w_{\rm{obs}}$&S/N&$\lambda_{\rm{obs}}$&Ident&z \\
 \hline
1&0.28& 58& 4064.12& ...& ...& & & & & \\ 
2&0.42& 44& 4152.90& SiIV\_1393 & 1.9796&$<$0.74 & && & \\ 
3&0.24& 41& 4178.74& SiIV\_1402 & 1.9789&$<$0.76 & && & \\ 
4&$<$0.34 & & & & &5.30& 50& 4236.43& NV\_1238   & 2.4197\\ 
5&0.97& 34& 4244.90& SiIV\_1393 & 2.0456&$<$3.32 & && & \\ 
6&$<$0.35 & & & & &3.95& 49& 4249.23& NV\_1242   & 2.4191\\
\hline 
\end{tabular}
 \end{center}
 \end{tiny}
\end{table}
\begin{table}
\caption{Line list for J~123510.5-010746 and J~123511.0-010830}
 \begin{tiny}
 \begin{center}
 \begin{tabular}{|p{0.0005cm} p{0.2cm} p{0.1cm} p{0.4cm} p{0.7cm} p{0.33cm} |  p{0.2cm} p{0.1cm} p{0.4cm} p{0.7cm} p{0.33cm}|}
 \hline
 & \multicolumn{3}{c}{J~123510.5-010746} & & & & \multicolumn{3}{c}{J~123511.0-010830}&\\
 &$w_{\rm{obs}}$&S/N&$\lambda_{\rm{obs}}$&Ident&z&$w_{\rm{obs}}$&S/N&$\lambda_{\rm{obs}}$&Ident&z \\
 \hline
1&$<$0.48 & & & & &1.44& 15& 3627.43& SiIV\_1393 & 1.6026\\ 
2&$<$0.42 & & & & &1.42& 18& 3650.60& SiIV\_1402 & 1.6024\\ 
3&$<$0.24 & & & & &1.08& 50& 3946.92& CIV\_1548  & 1.5494\\ 
4&$<$1.44 & & & & &0.78& 45& 3953.77& CIV\_1550  & 1.5495\\ 
5& & & & & &1.00& 42& 3974.10& ...& ...\\ 
6&$<$0.24 & & & & &2.57& 35& 4029.43& CIV\_1548  & 1.6027\\ 
7&$<$0.30 & & & & &2.34& 34& 4036.12& CIV\_1550  & 1.6026\\ 
8& & & & & &2.99& 32& 4193.30& ...& ...\\ 
9& & & & & &0.37& 37& 4318.36& ...& ...\\ 
10&$<$1.28 & & & & &1.34& 32& 4348.96& AlII\_1670 & 1.6029\\ 
11&$<$0.19 & & & & &2.23& 32& 4797.41& CIV\_1548  & 2.0987\\ 
12&$<$0.20 & & & & &1.64& 34& 4805.60& CIV\_1550  & 2.0988\\ 
13&$<$0.21 & & & & &0.94& 36& 4827.95& AlIII\_1854 & 1.6031\\ 
14&$<$0.21 & & & & &0.63& 38& 4848.76& AlIII\_1862 & 1.6030\\ 
15&0.24& 57& 4889.18& ...& ...& & & & & \\ 
16&1.12& 53& 5015.39& CIV\_1548  & 2.2395&$<$0.23 & && & \\ 
17&0.60& 54& 5023.48& CIV\_1550  & 2.2393&$<$0.24 & && & \\ 
18&0.29& 58& 5080.15& CIV\_1548  & 2.2813&$<$0.31 & && & \\ 
19&0.30& 57& 5088.41& CIV\_1550  & 2.2812&$<$0.32 & && & \\ 
20&0.29& 56& 5119.20& ...& ...& & & & & \\ 
21&0.23& 57& 5132.99& ...& ...& & & & & \\ 
 \hline 
\end{tabular}
 \end{center}
 \end{tiny}
\end{table}

\begin{table}
\caption{Line list for J~125556.9+001848 and J~125606.3+001728}
 \begin{tiny}
 \begin{center}
 \begin{tabular}{|p{0.0005cm} p{0.2cm} p{0.1cm} p{0.4cm} p{0.7cm} p{0.33cm} |  p{0.2cm} p{0.1cm} p{0.4cm} p{0.7cm} p{0.33cm}|}
 \hline
 & \multicolumn{3}{c}{J~125556.9+001848} & & & & \multicolumn{3}{c}{J~125606.3+001728}&\\
 &$w_{\rm{obs}}$&S/N&$\lambda_{\rm{obs}}$&Ident&z&$w_{\rm{obs}}$&S/N&$\lambda_{\rm{obs}}$&Ident&z \\
 \hline
1& & & & & &0.52& 35& 3807.24& ...& ...\\ 
2&0.53& 32& 3828.92& ...& ...& & & & & \\ 
3&$<$0.31 & & & & &1.01& 31& 3862.21& SiII\_1526 & 1.5298\\ 
4&$<$0.40 & & & & &2.02& 27& 3916.36& CIV\_1548  & 1.5296\\ 
5&$<$0.40 & & & & &1.78& 27& 3922.86& CIV\_1550  & 1.5296\\ 
6&$<$0.41 & & & & &1.00& 30& 4030.35& CII\_1334  & 2.0200\\
\hline 
 \end{tabular}
 \end{center}
 \end{tiny}
\end{table}
\clearpage
\begin{table}
\caption{Line list for J~135001.7-011703 and J~135003.0-011819}
 \begin{tiny}
 \begin{center}
 \begin{tabular}{|p{0.0005cm} p{0.2cm} p{0.1cm} p{0.4cm} p{0.7cm} p{0.33cm} |  p{0.2cm} p{0.1cm} p{0.4cm} p{0.7cm} p{0.33cm}|}
 \hline
 & \multicolumn{3}{c}{J~135001.7-011703} & & & & \multicolumn{3}{c}{J~135003.0-011819}&\\
 &$w_{\rm{obs}}$&S/N&$\lambda_{\rm{obs}}$&Ident&z&$w_{\rm{obs}}$&S/N&$\lambda_{\rm{obs}}$&Ident&z \\
 \hline
1&2.28& 51& 3863.44& MgII\_2796 & 0.3816&$<$0.87 & && & \\ 
2&1.04& 56& 3872.57& MgII\_2803 & 0.3813&$<$0.93 & && & \\ 
3&0.40& 43& 3908.55& ...& ...& & & & & \\ 
4&0.47& 43& 3937.48& NV\_1238   & 2.1784&$<$0.80 & && & \\ 
5&0.37& 45& 3950.47& NV\_1242   & 2.1787&$<$2.36 & && & \\ 
6&$<$0.31 & & & & &4.13& 11& 3969.05& SiII\_1190 & 2.3342\\
\hline 
\end{tabular}
 \end{center}
 \end{tiny}
\end{table}
\clearpage
\begin{table}
\caption{Line list for J~141124.6-022943 and J~141117.3-023222}
 \begin{tiny}
 \begin{center}
 \begin{tabular}{|p{0.0005cm} p{0.2cm} p{0.1cm} p{0.4cm} p{0.7cm} p{0.33cm} |  p{0.2cm} p{0.1cm} p{0.4cm} p{0.7cm} p{0.33cm}|}
 \hline
 & \multicolumn{3}{c}{J~141124.6-022943} & & & & \multicolumn{3}{c}{J~141117.3-023222}&\\
 &$w_{\rm{obs}}$&S/N&$\lambda_{\rm{obs}}$&Ident&z&$w_{\rm{obs}}$&S/N&$\lambda_{\rm{obs}}$&Ident&z \\
 \hline
1&$<$2.11 & & & & &0.87& 65& 4104.61& NV\_1238   & 2.3133\\ 
2&$<$6.53 & & & & &1.18& 63& 4117.01& NV\_1242   & 2.3127\\ 
3&$<$6.53 & & & & &1.17& 64& 4117.01& SiIV\_1393 & 1.9539\\ 
4&$<$0.29 & & & & &0.52& 58& 4142.86& SiIV\_1402 & 1.9533\\ 
5&$<$1.33 & & & & &1.25& 48& 4386.90& SiIV\_1393 & 2.1475\\ 
6&$<$0.86 & & & & &0.87& 50& 4415.04& SiIV\_1402 & 2.1474\\
\hline 
 \end{tabular}
 \end{center}
 \end{tiny}
\end{table}

\begin{table}
\caption{Line list for J~144245.6-024251 and J~144245.7-023906}
 \begin{tiny}
 \begin{center}
 \begin{tabular}{|p{0.0005cm} p{0.2cm} p{0.1cm} p{0.4cm} p{0.7cm} p{0.33cm} |  p{0.2cm} p{0.1cm} p{0.4cm} p{0.7cm} p{0.33cm}|}
 \hline
 & \multicolumn{3}{c}{J~144245.6-024251} & & & & \multicolumn{3}{c}{J~144245.7-023906}&\\
 &$w_{\rm{obs}}$&S/N&$\lambda_{\rm{obs}}$&Ident&z&$w_{\rm{obs}}$&S/N&$\lambda_{\rm{obs}}$&Ident&z \\
 \hline
1&1.30& 21& 3830.49& SiII\_1190 & 2.2178&$<$0.51 & && & \\ 
2&1.45& 20& 3840.14& SiII\_1193 & 2.2181&$<$0.52 & && & \\ 
3&4.24& 18& 3881.79& SiIII\_1206 & 2.2174&$<$0.51 & && & \\ 
& & & & HI\_1215   & 2.193 & & && & \\ 
4&2.99& 47& 4055.69& SiII\_1260 & 2.2177&$<$1.73 & && & \\ 
& & & & HI\_1215   & 2.336 & & && & \\
\hline 
\end{tabular}
 \end{center}
 \end{tiny}
\end{table}

\begin{table}
\caption{Line list for Q2129-4653A and Q2129-4653B}
 \begin{tiny}
 \begin{center}
 \begin{tabular}{|p{0.0005cm} p{0.2cm} p{0.1cm} p{0.4cm} p{0.7cm} p{0.33cm} |  p{0.2cm} p{0.1cm} p{0.4cm} p{0.7cm} p{0.33cm}|}
 \hline
 & \multicolumn{3}{c}{Q2129-4653A} & & & & \multicolumn{3}{c}{Q2129-4653B}&\\
 &$w_{\rm{obs}}$&S/N&$\lambda_{\rm{obs}}$&Ident&z&$w_{\rm{obs}}$&S/N&$\lambda_{\rm{obs}}$&Ident&z \\
 \hline
1&0.42& 39& 3916.03& SiIV\_1393 & 1.8097&$<$1.24 & && & \\ 
2&0.25& 34& 3933.37& ...& ...& & & & & \\ 
3&0.33& 34& 3941.81& SiIV\_1402 & 1.8100&$<$0.23 & && & \\ 
4&0.31& 34& 3990.91& SiII\_1526 & 1.6141&$<$0.27 & && & \\ 
5&$<$0.94 & & & & &0.51& 36& 4044.41& CIV\_1548  & 1.6123\\ 
6&1.36& 24& 4047.24& CIV\_1548  & 1.6142&$<$0.45 & && & \\ 
7&1.01& 24& 4051.70& CIV\_1550  & 1.6127&$<$0.45 & && & \\ 
8&$<$1.52 & & & & &1.21& 33& 4051.84& CIV\_1550  & 1.6128\\ 
& & & & & & & &  & CIV\_1548  & 1.6171\\ 
9&0.65& 23& 4052.14& CIV\_1548  & 1.6173&$<$0.42 & && & \\ 
10&0.28& 26& 4055.71& CIV\_1550  & 1.6153&$<$0.38 & && & \\ 
11&$<$0.50 & & & & &0.56& 34& 4058.62& CIV\_1550  & 1.6171\\ 
12&1.14& 23& 4186.23& CIV\_1548  & 1.7039&$<$0.40 & && & \\ 
13&0.62& 24& 4193.46& CIV\_1550  & 1.7041&$<$0.39 & && & \\ 
14&0.51& 24& 4290.52& SiII\_1526 & 1.8103&$<$0.39 & && & \\ 
15&$<$0.49 & & & & &0.83& 31& 4327.13& CIV\_1548  & 1.7949\\ 
16&$<$0.49 & & & & &0.75& 32& 4334.42& CIV\_1550  & 1.7950\\ 
17&1.26& 23& 4350.04& CIV\_1548  & 1.8097&$<$0.38 & && & \\ 
18&0.82& 23& 4357.06& CIV\_1550  & 1.8096&$<$0.39 & && & \\ 
19&0.40& 24& 4367.63& AlII\_1670 & 1.6141&$<$0.39 & && & \\ 
20&0.74& 25& 4439.61& ...& ...& & & & & \\ 
21&0.35& 25& 4694.68& AlII\_1670 & 1.8099&$<$0.39 & && & \\ 
28&0.89& 22& 5526.40& MgII\_2796 & 0.9763&$<$0.40 & && & \\ 
29&0.61& 23& 5540.46& MgII\_2803 & 0.9762&$<$0.40 & && & \\ 
 \hline 
 \end{tabular}
 \end{center}
 \end{tiny}
\end{table}

\begin{table}
\caption{Line list for Q2139-4504A and Q2139-4504B}
 \begin{tiny}
 \begin{center}
 \begin{tabular}{|p{0.0005cm} p{0.2cm} p{0.1cm} p{0.4cm} p{0.7cm} p{0.33cm} |  p{0.2cm} p{0.1cm} p{0.4cm} p{0.7cm} p{0.33cm}|}
 \hline
 & \multicolumn{3}{c}{Q2139-4504A} & & & & \multicolumn{3}{c}{Q2139-4504B}&\\
 &$w_{\rm{obs}}$&S/N&$\lambda_{\rm{obs}}$&Ident&z&$w_{\rm{obs}}$&S/N&$\lambda_{\rm{obs}}$&Ident&z \\
 \hline
1&$<$2.25 & & & & &8.02& 12& 4320.48& CII\_1036  & 3.1690\\ 
2&$<$0.34 & & & & &8.29& 12& 4962.75& SiII\_1190 & 3.1689\\ 
& & & & & & & &  & HI\_1215   & 3.0823\\ 
3&$<$0.40 & & & & &5.51& 16& 4973.07& SiII\_1193 & 3.1675\\ 
& & & & & & & &  & HI\_1215   & 3.0908\\ 
6&1.59& 31& 5001.80& SiIV\_1393: & 2.5887&$<$4.93 & && & \\ 
7&0.80& 34& 5024.51& SiIV\_1402: & 2.5818&$<$6.61 & && & \\ 
9&$<$0.35 & & & & &3.19& 16& 5027.95& SiIII\_1206 & 3.1674\\ 
10&0.82& 20& 5130.70& ...& ...& & & & & \\ 
11& & & & & &0.34& 46& 5224.46& ...& ...\\ 
12& & & & & &2.33& 40& 5234.87& ...& ...\\ 
13&0.83& 19& 5240.81& ...& ...& & & & & \\ 
14&$<$0.69 & & & & &8.75& 26& 5253.21& SiII\_1260 & 3.1678\\ 
15&0.81& 21& 5286.11& ...& ...& & & & & \\ 
16&$<$0.65 & & & & &0.87& 34& 5355.71& SiII\_1260 & 3.2491\\ 
20&$<$0.67 & & & & &6.52& 18& 5427.63& OI\_1302   & 3.1681\\ 
21&$<$0.68 & & & & &3.44& 24& 5436.45& SiII\_1304 & 3.1679\\ 
25&1.19& 17& 5557.65& CIV\_1548  & 2.5897&$<$4.26 & && & \\ 
26&$<$2.11 & & & & &8.94& 19& 5562.47& CII\_1334  & 3.1681\\ 
& & & & & & & &  & HI\_1215   & 3.5756\\ 
27&1.01& 18& 5567.12& CIV\_1550  & 2.5899&$<$3.55 & && & \\ 
28&1.20& 20& 5604.82& ...& ...& & & & & \\ 
29&$<$0.55 & & & & &1.04& 28& 5670.52& CII\_1334  & 3.2491\\ 
30&0.74& 21& 5688.60& ...& ...& & & & & \\ 
31&0.95& 20& 5704.82& ...& ...& & & & & \\ 
 \hline 
\end{tabular}
 \end{center}
 \end{tiny}
\end{table}

\begin{table}
\caption{Line list for J~214501.6-303121 and J~214507.0-303046}
 \begin{tiny}
 \begin{center}
 \begin{tabular}{|p{0.0005cm} p{0.2cm} p{0.1cm} p{0.4cm} p{0.7cm} p{0.33cm} |  p{0.2cm} p{0.1cm} p{0.4cm} p{0.7cm} p{0.33cm}|}
 \hline
 & \multicolumn{3}{c}{J~214501.6-303121} & & & & \multicolumn{3}{c}{J~214507.0-303046}&\\
 &$w_{\rm{obs}}$&S/N&$\lambda_{\rm{obs}}$&Ident&z&$w_{\rm{obs}}$&S/N&$\lambda_{\rm{obs}}$&Ident&z \\
 \hline
1&$<$0.61 & & & & &3.99& 16& 3641.56& OVI\_1031  & 2.5289\\ 
2&$<$0.57 & & & & &4.03& 23& 3661.77& OVI\_1037  & 2.5290\\ 
3&0.36& 57& 3956.66& ...& ...& & & & & \\ 
4&0.60& 56& 3968.16& ...& ...& & & & & \\ 
5&0.23& 55& 3980.43& ...& ...& & & & & \\ 
6&0.33& 55& 3991.35& ...& ...& & & & & \\ 
7&0.44& 51& 4002.21& ...& ...& & & & & \\ 
8&$<$0.28 & & & & &1.37& 39& 4087.01& SiIV\_1393 & 1.9324\\ 
& & & & & & & &  & HI\_1215   & 2.3619\\ 
9&$<$0.29 & & & & &0.43& 42& 4113.24& SiIV\_1402 & 1.9322\\ 
& & & & & & & &  & HI\_1215   & 2.3835\\ 
10&0.25& 42& 4204.39& CIV\_1548  & 1.7157&$<$0.26 & && & \\ 
11&0.26& 43& 4210.65& CIV\_1550  & 1.7152&$<$0.45 & && & \\ 
12&0.45& 42& 4305.48& ...& ...& & & & & \\ 
13&$<$0.28 & & & & &3.62& 57& 4371.73& NV\_1238   & 2.5289\\ 
14&$<$0.28 & & & & &2.65& 60& 4385.68& NV\_1242   & 2.5289\\ 
16&0.21& 47& 4464.60& SiIV\_1393: & 2.2033&$<$0.23 & && & \\ 
17&0.11& 49& 4494.86& SiIV\_1402 & 2.2043&$<$0.24 & && & \\ 
18&$<$0.26 & & & & &3.90& 40& 4540.16& CIV\_1548  & 1.9325\\ 
19&$<$0.26 & & & & &2.37& 41& 4547.69& CIV\_1550  & 1.9325\\ 
20&$<$0.24 & & & & &0.98& 51& 4888.98& CIV\_1548  & 2.1578\\ 
21&$<$0.24 & & & & &0.57& 53& 4897.20& CIV\_1550  & 2.1579\\ 
22&0.25& 52& 4901.73& ...& ...& & & & & \\ 
23&$<$0.22 & & & & &0.50& 56& 4917.76& SiIV\_1393 & 2.5284\\ 
24&$<$0.20 & & & & &0.25& 58& 4949.59& SiIV\_1402 & 2.5284\\ 
25&1.46& 60& 4961.46& CIV\_1548  & 2.2047&$<$0.21 & && & \\ 
26&0.75& 63& 4969.77& CIV\_1550: & 2.2047&$<$0.21 & && & \\ 
32&$<$0.28 & & & & &2.72& 56& 5463.43& CIV\_1548  & 2.5289\\ 
34&$<$0.28 & & & & &2.92& 60& 5471.99& CIV\_1550  & 2.5285\\ 
 \hline 
\end{tabular}
 \end{center}
 \end{tiny}
\end{table}

\begin{table}
\caption{Line list for J~215225.8-283058 and J~215240.0-283251}
 \begin{tiny}
 \begin{center}
 \begin{tabular}{|p{0.0005cm} p{0.2cm} p{0.1cm} p{0.4cm} p{0.7cm} p{0.33cm} |  p{0.2cm} p{0.1cm} p{0.4cm} p{0.7cm} p{0.33cm}|}
 \hline
 & \multicolumn{3}{c}{J~215225.8-283058} & & & & \multicolumn{3}{c}{J~215240.0-283251}&\\
 &$w_{\rm{obs}}$&S/N&$\lambda_{\rm{obs}}$&Ident&z&$w_{\rm{obs}}$&S/N&$\lambda_{\rm{obs}}$&Ident&z \\
 \hline
1&$<$3.33 & & & & &2.20& 22& 3817.99& OVI\_1031: & 2.6999\\ 
2&$<$0.72 & & & & &1.33& 27& 3839.42& OVI\_1037: & 2.7002\\ 
3&$<$2.08 & & & & &2.55& 38& 4523.59& SiIV\_1393 & 2.2456\\ 
4&$<$0.15 & & & & &1.00& 41& 4552.89& SiIV\_1402 & 2.2456\\ 
5&0.30& 59& 4579.00& ...& ...& & & & & \\ 
6&$<$0.33 & & & & &0.83& 41& 4582.40& CIV\_1548  & 1.9598\\
\hline 
 \end{tabular}
 \end{center}
 \end{tiny}
\end{table}

\begin{table}
\caption{Line list for J~223850.1-295612 and J~223850.9-295301}
 \begin{tiny}
 \begin{center}
 \begin{tabular}{|p{0.0005cm} p{0.2cm} p{0.1cm} p{0.4cm} p{0.7cm} p{0.33cm} |  p{0.2cm} p{0.1cm} p{0.4cm} p{0.7cm} p{0.33cm}|}
 \hline
 & \multicolumn{3}{c}{J~223850.1-295612} & & & & \multicolumn{3}{c}{J~223850.9-295301}&\\
 &$w_{\rm{obs}}$&S/N&$\lambda_{\rm{obs}}$&Ident&z&$w_{\rm{obs}}$&S/N&$\lambda_{\rm{obs}}$&Ident&z \\
 \hline
1&0.62& 62& 4270.70& ...& ...& & & & & \\ 
2&1.47& 57& 4293.80& ...& ...& & & & & \\ 
3&0.38& 49& 4343.59& ...& ...& & & & & \\ 
5&0.53& 46& 4348.16& ...& ...& & & & & \\ 
6&0.79& 45& 4364.05& CIV\_1548  & 1.8188&$<$0.26 & && & \\ 
7&0.62& 44& 4370.17& CIV\_1550  & 1.8180&$<$0.26 & && & \\
\hline 
\end{tabular}
 \end{center}
 \end{tiny}
\end{table}

\begin{table}
\caption{Line list for J~230301.6-290027 and J~230318.4-290120}
 \begin{tiny}
 \begin{center}
 \begin{tabular}{|p{0.0005cm} p{0.2cm} p{0.1cm} p{0.4cm} p{0.7cm} p{0.33cm} |  p{0.2cm} p{0.1cm} p{0.4cm} p{0.7cm} p{0.33cm}|}
 \hline
 & \multicolumn{3}{c}{J~230301.6-290027} & & & & \multicolumn{3}{c}{J~230318.4-290120}&\\
 &$w_{\rm{obs}}$&S/N&$\lambda_{\rm{obs}}$&Ident&z&$w_{\rm{obs}}$&S/N&$\lambda_{\rm{obs}}$&Ident&z \\
 \hline
1&0.97& 47& 4430.38& CIV\_1548  & 1.8616&$<$0.21 & && & \\ 
2&0.55& 46& 4438.40& CIV\_1550  & 1.8620&$<$0.21 & && & \\ 
3&$<$0.27 & & & & &0.43& 57& 4444.72& NV\_1238   & 2.5879\\ 
4&$<$0.29 & & & & &0.33& 57& 4459.40& NV\_1242   & 2.5882\\ 
5&0.52& 43& 4464.84& CIV\_1548  & 1.8839&$<$0.22 & && & \\ 
6&0.26& 42& 4472.66& CIV\_1550  & 1.8841&$<$0.23 & && & \\ 
7&0.18& 33& 4802.82& ...& ...& & & & & \\ 
9&0.49& 34& 4908.45& ...& ...& & & & & \\ 
10&$<$0.32 & & & & &0.68& 42& 4973.79& FeII\_2344 & 1.1217\\ 
11&$<$0.34 & & & & &1.07& 41& 5055.85& FeII\_2382 & 1.1218\\ 
12&0.41& 34& 5069.16& CIV\_1548: & 2.2742&$<$0.29 & && & \\ 
13&0.52& 34& 5077.52& CIV\_1550: & 2.2742&$<$0.29 & && & \\ 
14&$<$0.36 & & & & &0.46& 39& 5156.94& MgII\_2796 & 0.8442\\ 
15&$<$0.36 & & & & &0.55& 39& 5170.83& MgII\_2803 & 0.8444\\ 
16&1.32& 32& 5267.46& CIV\_1548  & 2.4023&$<$0.31 & && & \\ 
17&0.78& 32& 5276.16& CIV\_1550  & 2.4023&$<$0.30 & && & \\ 
18& & & & & &0.22& 40& 5319.04& ...& ...\\ 
19&0.30& 34& 5363.27& ...& ...& & & & & \\ 
20&$<$0.27 & & & & &0.57& 50& 5488.50& FeII\_2586 & 1.1219\\ 
21&$<$0.26 & & & & &0.97& 54& 5516.90& FeII\_2600 & 1.1217\\ 
 \hline 
 \end{tabular}
 \end{center}
 \end{tiny}
 \end{table}
\clearpage

\section{Spectra}
 \begin{figure*}
 \unitlength=1cm
 \vbox to220mm{\vfil{\psfig{height=22.5cm,width=18.5cm,angle=-90,figure=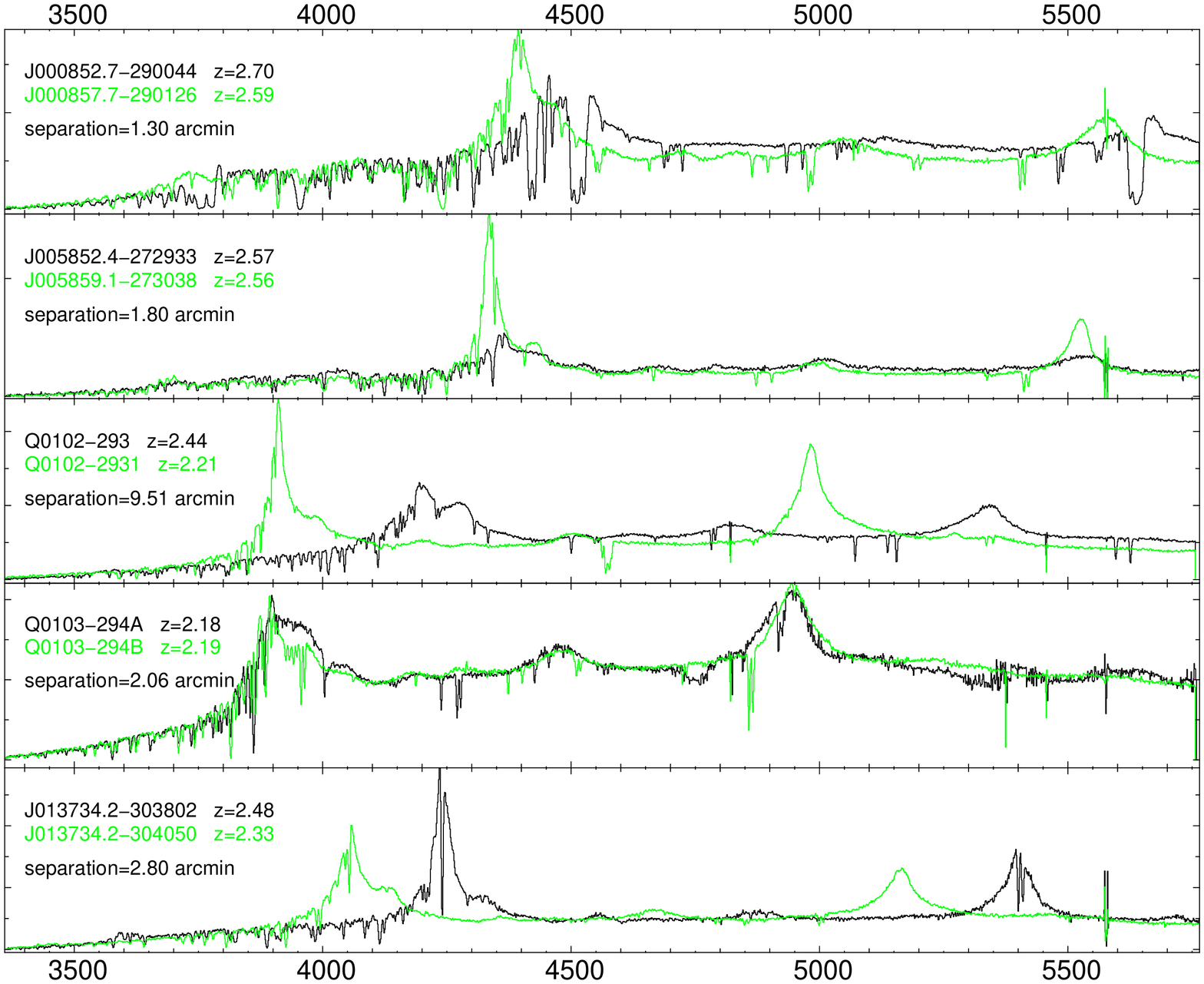 }}
 \caption{Spectra of the observed QSO pairs in order of increasing right ascencion.
 The names of the QSOs, their emission redshift and the separation between the
 two lines of sight are indicating in the top-left corner of each sub-panel.
 }
\label{f:spectra}
\vfil}
 \end{figure*}
 \begin{figure*}
 \unitlength=1cm
 \vbox to220mm{\vfil{\psfig{height=22.5cm,width=18.5cm,angle=-90,figure=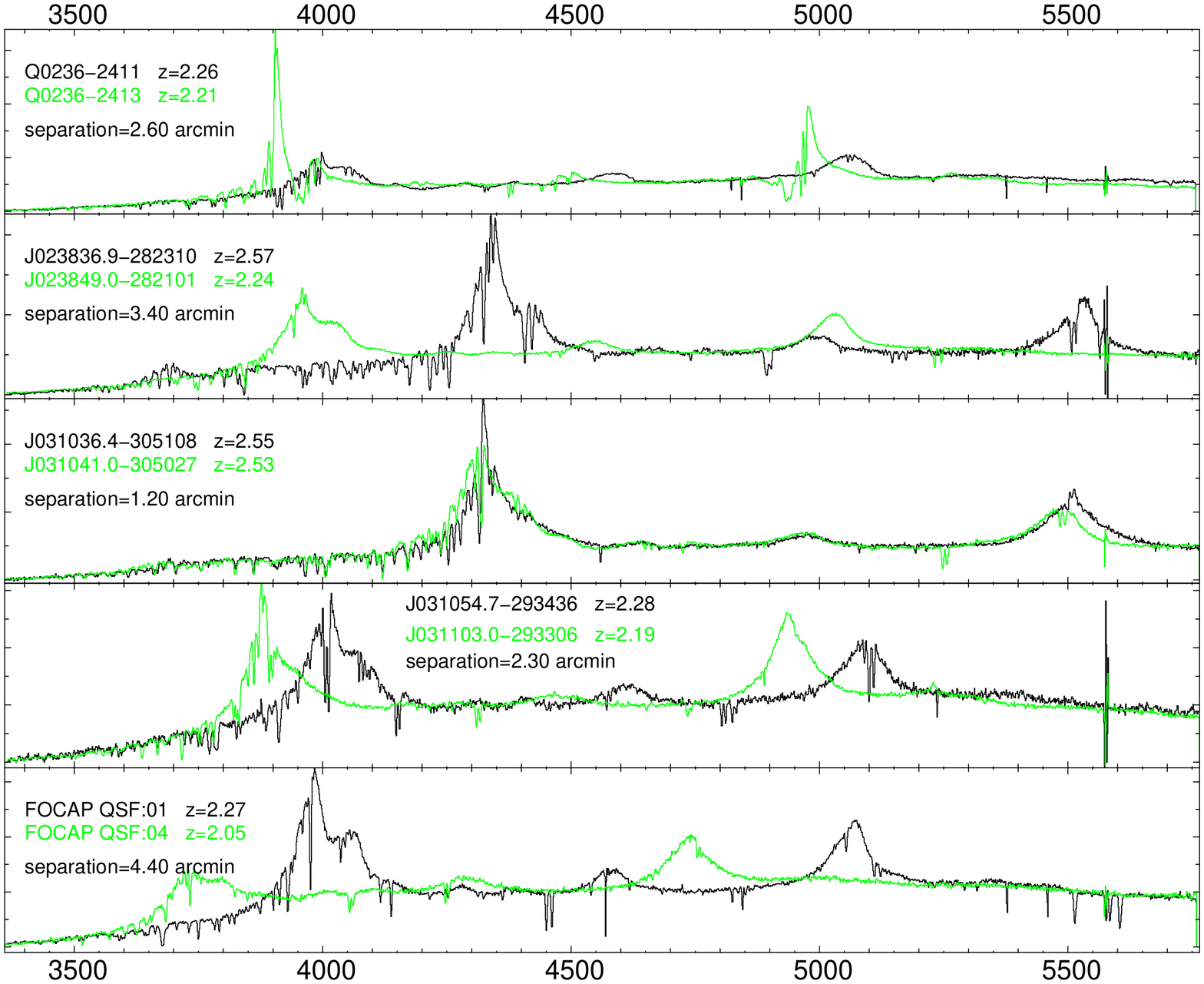 }}
 \caption{As in Figure C1.}
\vfil}
 \end{figure*}
 \begin{figure*}
 \unitlength=1cm
 \vbox to220mm{\vfil{\psfig{height=22.5cm,width=18.5cm,angle=-90,figure=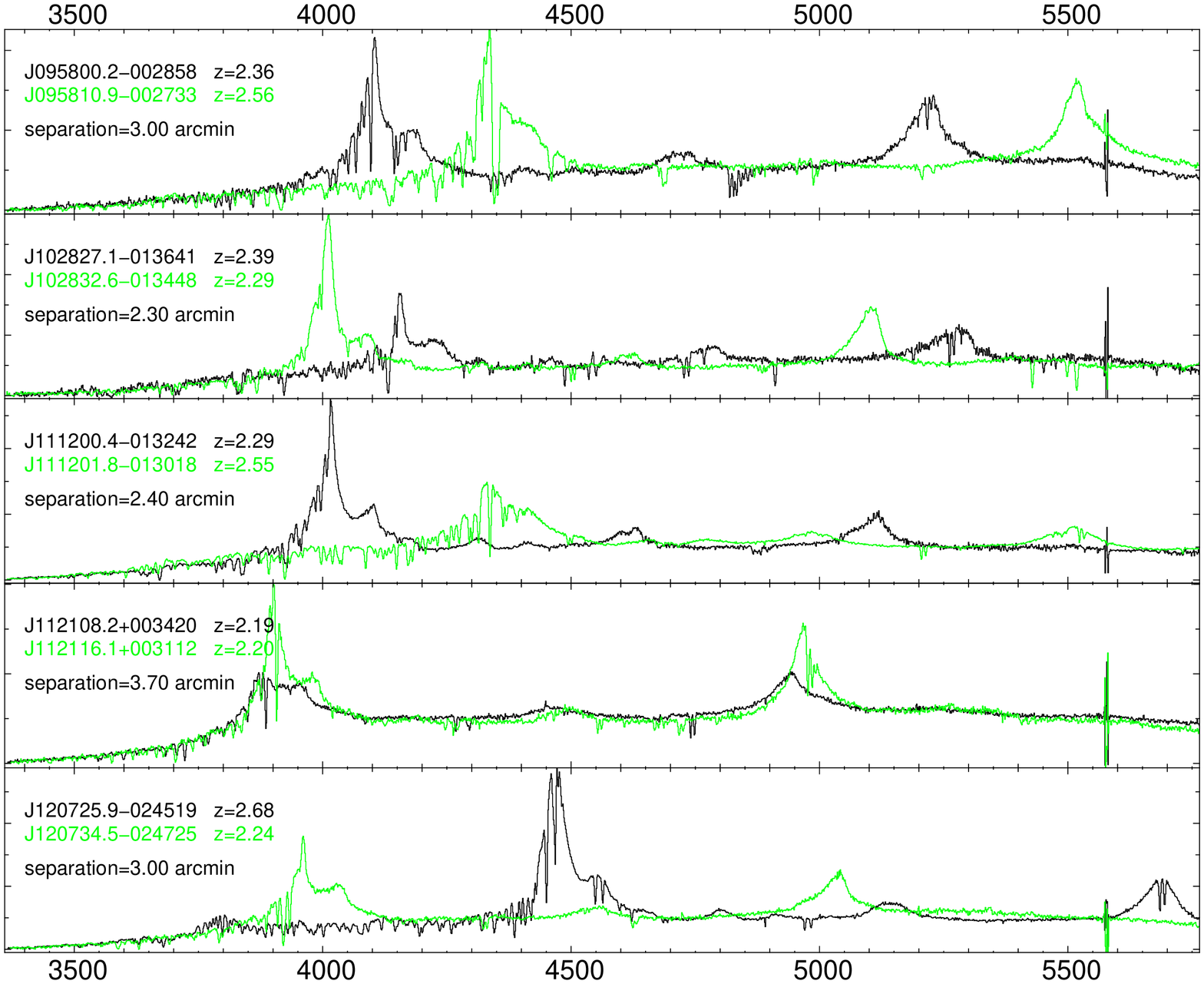 }}
 \caption{As in Figure C1.}
\vfil}
 \end{figure*}
 \begin{figure*}
 \unitlength=1cm
 \vbox to220mm{\vfil{\psfig{height=22.5cm,width=18.5cm,angle=-90,figure=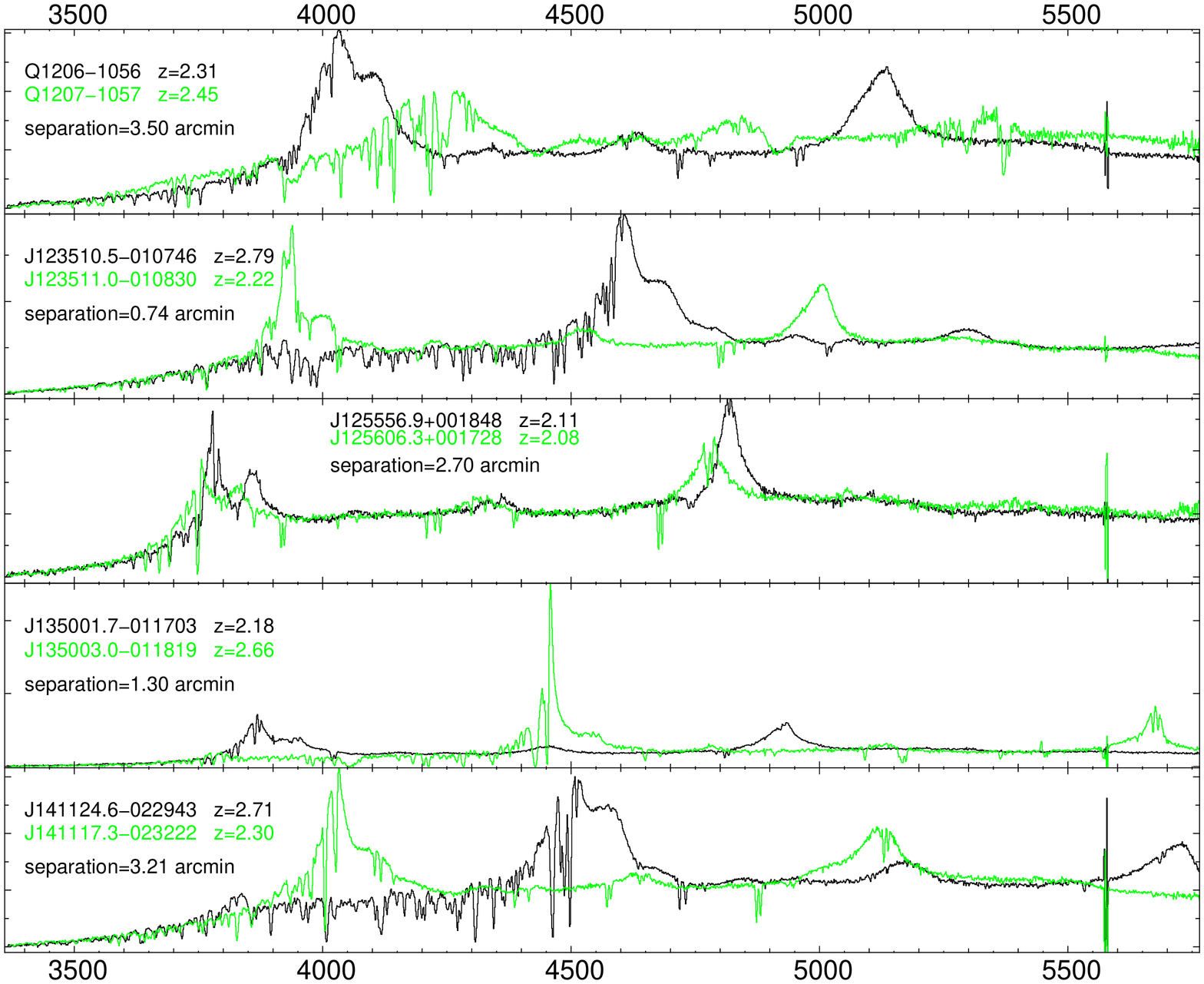 }}
 \caption{As in Figure C1.}
\vfil}
 \end{figure*}
 \begin{figure*}
 \unitlength=1cm
 \vbox to220mm{\vfil{\psfig{height=22.5cm,width=18.5cm,angle=-90,figure=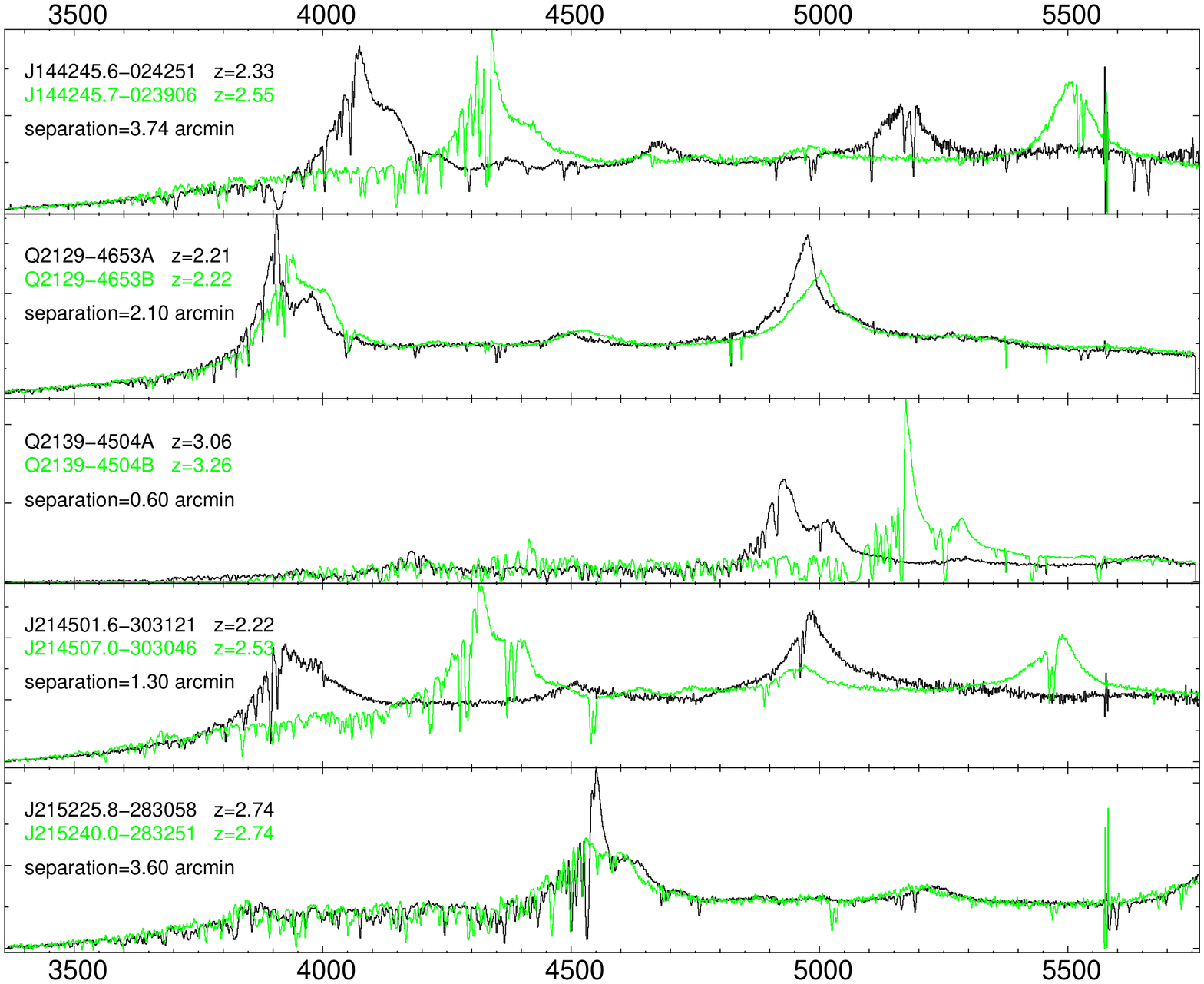 }}
 \caption{As in Figure C1.}
\vfil}
 \end{figure*}
 \begin{figure*}
 \unitlength=1cm
 \vbox to144mm{\vfil{\psfig{height=14.4cm,width=18.5cm,angle=-90,figure=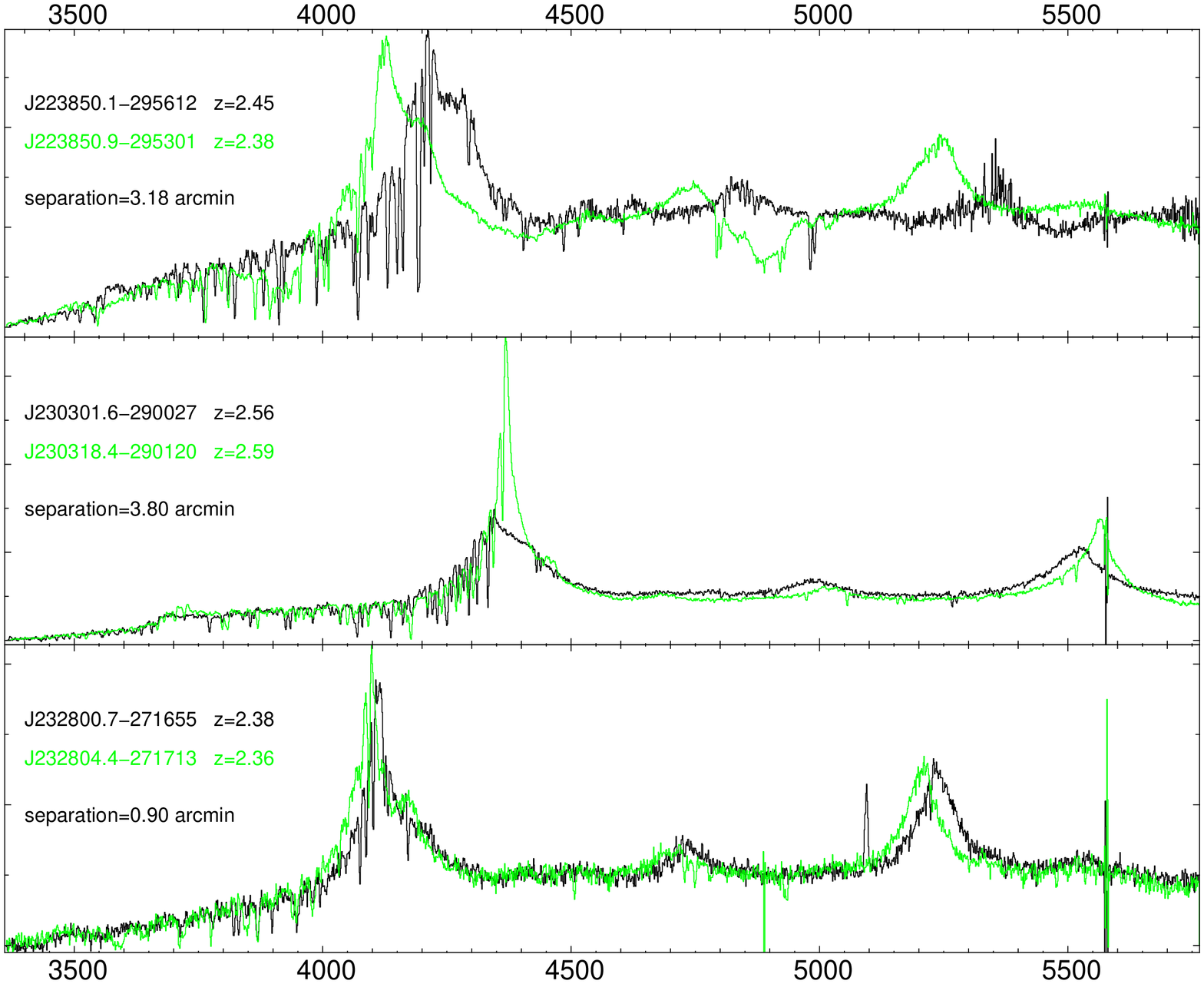 }}
\caption{As in Figure C1.} 
\vfil}
 \end{figure*}

 \end{document}